\documentclass[final,twocolumn]{IEEEtran}

\usepackage{graphicx}
\usepackage{amssymb}
 
\usepackage{amsmath}
\usepackage{dsfont}
\usepackage{bm}
\usepackage{subfigure}

\def\C{{\mathds C}}

\newcommand{\be}{\begin{equation}}
\newcommand{\ee}{\end{equation}}

\newcommand{\bzero}{{\mbox{\boldmath $0$}}}
\newcommand{\bI}{{\mbox{\boldmath $I$}}}
\newcommand{\bz}{{\mbox{\boldmath $z$}}}
\newcommand{\bn}{{\mbox{\boldmath $n$}}}
\newcommand{\bv}{{\mbox{\boldmath $v$}}}
\newcommand{\bp}{{\mbox{\boldmath $p$}}}
\newcommand{\bu}{{\mbox{\boldmath $u$}}}
\newcommand{\bx}{{\mbox{\boldmath $x$}}}
\newcommand{\bor}{{\mbox{\boldmath $r$}}}

\newcommand{\bC}{{\mbox{\boldmath $C$}}}

\newcommand{\bX}{{\mbox{\boldmath $X$}}}

\newcommand{\bR}{{\mbox{\boldmath $R$}}}

\newcommand{\bS}{{\mbox{\boldmath $S$}}}
\newcommand{\bP}{{\mbox{\boldmath $P$}}}

\newcommand{\bSigma}{{\mbox{\boldmath $\Sigma$}}}

\newcommand{\Puperp}{{\mbox{\boldmath $\bP^\perp_{\bu}$}}}

\newcommand{\Putilde}{{\mbox{\boldmath $\bP_{\!\tilde{\bu}}$}}}
\newcommand{\Putildeperp}{{\mbox{\boldmath $\bP^\perp_{\tilde{\bu}}$}}}

\newcommand{\Pcomplex}{{\mbox{\boldmath $\bP^\perp_{\bP^\perp_{\tilde{\bu}}\tilde{\bv}}$}}}

\newcommand{\Pvtilde}{{\mbox{\boldmath $\bP_{\!\tilde{\bv}}$}}}
\newcommand{\Pvtildeperp}{{\mbox{\boldmath $\bP^\perp_{\tilde{\bv}}$}}}

\newtheorem{proposition}{Proposition}

\newcommand{\test}{\mbox{$
\begin{array}{c}
\stackrel{ \stackrel{\textstyle H_1}{\textstyle >} }{ 
\stackrel{\textstyle <}{ \textstyle  H_0} }

\end{array}
$}}

\usepackage{color}

\begin{document}

\title{
Design of robust radar detectors through random perturbation of the target signature 
}

\author{Angelo Coluccia, \IEEEmembership{Senior Member, IEEE}, Giuseppe Ricci, \IEEEmembership{Senior Member, IEEE}, and Olivier Besson  
\thanks{A. Coluccia and G. Ricci are with the Dipartimento di Ingegneria dell'Innovazione,
        Universit\`a del Salento, Via Monteroni, 73100 Lecce, Italy.
        E-Mail: angelo.coluccia@unisalento.it, giuseppe.ricci@unisalento.it.}
\thanks{O. Besson is with ISAE-SUPAERO, 10 Avenue Edouard Belin, 31055 Toulouse, France. E-mail: olivier.besson@isae-supaero.fr.}         
}

\markboth{\copyright IEEE Transactions on Signal Processing,  Vol. 67, Issue 19, Oct. 2019, DOI: 10.1109/TSP.2019.2935915}{}

\maketitle

\begin{abstract}
The paper addresses the problem of designing radar detectors more robust than Kelly's detector to possible mismatches of the assumed target signature, but with no performance degradation under matched conditions. The idea is to model the received signal under the signal-plus-noise hypothesis by adding a random component, parameterized via a design covariance matrix, that makes the hypothesis more plausible in presence of mismatches. Moreover, an unknown power of such component, to be estimated from the observables, can lead to no performance loss. Derivation of the (one-step) GLRT is provided for two choices of the design matrix, obtaining detectors with different complexity and behavior. A third parametric detector is also obtained by an ad-hoc generalization of one of such GLRTs. The analysis shows that the proposed approach can cover a range of different robustness levels that is not achievable by state-of-the-art with the same performance of Kelly's detector under matched conditions.
\end{abstract}


\section{Introduction}

\subsection{Notation}

Vectors and matrices are denoted by boldface lower-case and upper-case letters, respectively.
The symbols $|\cdot|$, $\| \cdot \|$, $\det(\cdot)$, 
$\mbox{tr}(\cdot)$, ${}^T$,
${}^\dag$, denote modulus value, Euclidean norm, determinant, 
trace, transpose, and conjugate transpose (Hermitian), respectively.
$E[\cdot]$ is the statistical expectation operator. 
$\C$ is the set of complex numbers and $\C^{N\times M}$ is the Euclidean space of $(N\times M)$-dimensional complex matrices. 
$\bzero$ is the null vector of proper dimension and 
$\bI_N$ stands for the $N \times N$ identity matrix.
$\Puperp$ is the (orthogonal) projection matrix onto the orthogonal complement of the subspace spanned by the vector $\bu$.
Finally, we write $\bx\sim
{\cal CN}_N (\bzero, \bm{M})$ if $\bx$ is an $N$-dimensional complex normal vector with zero mean and (Hermitian) positive definite covariance matrix $\bm{M}$.

\subsection{Existing results and motivation}

The well-known  problem of detecting the possible presence of a coherent return from a given cell under test (CUT) in range, doppler, and azimuth, is classically formulated as the following
hypothesis testing problem:
\begin{equation}
\left\{
\begin{array}{ll}
H_{0}: & \bz =  \bn \\
H_{1}: & \bz = \alpha \bv + \bn
\end{array} 
\right.
\end{equation}
where
$\bz \in \C^{N \times 1}$, $\bn \in \C^{N \times 1}$, and $\bv \in \C^{N \times 1}$ are 
the received vector,  corresponding noise term, and 
 known space-time steering vector of the useful target echo.
In general $N$ is the number of processed samples from the CUT; it might be the number of antenna array elements times the number of pulses \cite{Ward,Klemm-STAP}.
The noise term is often modeled according to the complex normal distribution
with zero mean and unknown (Hermitian) positive definite matrix $\bC$,  i.e., $\bm{n} \sim {\cal CN}_N (\bzero, \bC)$.
Modeling $\alpha \in \C$ as an unknown deterministic parameter
returns a complex normal distribution for $\bz$ under both hypotheses; the non-zero mean under $H_1$ makes it possible to discriminate the two hypotheses
\begin{equation}
\left\{
\begin{array}{ll}
H_{0}: & \bz  \thicksim {\cal CN}_N (\bzero, \bC) \\ 
H_{1}: & \bz \thicksim {\cal CN}_N (\alpha \bv, \bC).
\end{array}
\right.
\label{eq:test1ord}
\end{equation}
In the pioneering paper by Kelly \cite{Kelly},  the
generalized likelihood ratio test (GLRT) is derived for  \eqref{eq:test1ord}, assuming a set of $K \geq N$ independent and identically distributed training (or secondary) data $\bor_1, \ldots, \bor_{K}$, independent also of $\bz$, free of target echoes, and  sharing with the CUT the statistical characteristics of the noise, is available.
 In  \cite{Kelly89} the performance of such a detector is assessed when the actual
steering vector is not aligned with the nominal one.
The analysis shows that it is a selective receiver, i.e., it may have excellent rejection capabilities of signals arriving from directions  different from the nominal one. A selective detector is desirable for target localization. Instead, when a radar is working in searching mode, a certain level of robustness to mismatches is preferable.
For this reason, many  works have addressed the problem of enhancing either the selectivity or the robustness of radar detectors to  mismatches. In particular, the adaptive matched filter (AMF) \cite{Kelly-Nitzberg}, which solves \eqref{eq:test1ord} following a two-step approach,  is a prominent example of robust detector, while the adaptive coherence estimator (ACE, also known as adaptive normalized 
matched filter)  \cite{ACE} is another example of selective receiver. Other relevant examples of selective receivers are obtained by solving a modified hypothesis testing problem that assumes the presence of a (fictitious) coherent signal under the noise-only ($H_0$) hypothesis to make it more plausible in presence of signal mismatches \cite{Pulsone-Rader,Fabrizio-Farina,W-ABORT}.
A family of receivers, obtained by inserting a nonnegative parameter in the original Kelly's detector, has been proposed  by Kalson in \cite{Kalson}; such a parameter, indicated as $\beta$ in the following,
allows one to control the degree to which mismatched signals are rejected,  so obtaining behaviors in between the AMF and Kelly's detector.
A different tunable receiver called KWA has been proposed in \cite{KWA}, which encompasses as special cases Kelly's  and  W-ABORT detectors \cite{W-ABORT}. Although designed for enhanced selectivity, for values of its tunable parameter $\gamma$ smaller than $1/2$ it behaves as a robust detector,  reaching the energy detector for $\gamma=0$.
Tunable receivers have also been proposed in \cite{JunLiu1,JunLiu2}.
Other approaches, as for instance those based 
on the cone idea, 
can guarantee an increased robustness at the price of a certain loss under matched conditions \cite{Coni-SOC,Besson1}.
 A robust two-stage detector obtained by cascading a GLRT-based subspace detector
and the Rao test has been proposed in \cite{Chenpeng}.
Robust and selective detectors have also been considered in the context of subspace detection \cite{JunLiu3,Liu,Duan}.
Multidimensional/multichannel signal detection in homogeneous or partially homogeneous Gaussian disturbance (with unknown covariance matrix and unknown structured deterministic interference) is considered in \cite{Orlando1,Orlando2}.

On the other hand, modeling $\alpha \in \C$ as a complex normal random variable with zero mean and variance $\overline{|\alpha|^2} = E[|\alpha|^2]$,
returns a zero-mean complex normal distribution for $\bz$ under both hypotheses; 
hence
\begin{equation}
\left\{
\begin{array}{ll}
H_{0}: & \bz  \thicksim {\cal CN}_N (\bzero, \bC) \\ 
H_{1}: & \bz \thicksim {\cal CN}_N (\bzero, \bC+ \overline{|\alpha|^2} \bv \bv^{\dagger} ).
\end{array}
\right.
\label{eq:test2ord}
\end{equation}
This is a ``second-order'' approach to target modeling, i.e., the presence of a useful signal ($H_1$ hypothesis) is modeled in terms of a  modification of the noise covariance matrix, instead of appearing in the mean as conversely for the more classical problem \eqref{eq:test1ord}. Interesting properties in terms of either rejection capabilities or robustness to mismatches on the nominal steering vector 
can be obtained by considering a random (instead of deterministic) target signal, depending on the way problem \eqref{eq:test2ord} is solved and possibly on the presence of a fictitious signal under $H_0$ \cite{CAMSAP,Besson_collaboration}. 

 In this paper,  we investigate the potential of a detector that solves 
 the following new hypothesis testing problem
\begin{equation}
\left\{
\begin{array}{ll}
H_{0}: & \bz =  \bn \\
H_{1}: & \bz = \alpha \bv+\bm{\theta} + \bn 
\end{array}
\right.\label{eq:test}
\end{equation}
where $\alpha$ is an unknown deterministic parameter and $\bm{\theta} \thicksim {\cal CN}_N (\bzero, \nu \bm{\Sigma})$ represents  a random   component. 
The goal is to obtain a detector that exhibits the same probability of detection ($P_d$) of Kelly's GLRT under matched conditions, but  is more robust than the latter to mismatches between the nominal steering vector and the actual one. 
In the existing literature, the ``win-win'' situation in which robustness is achieved without any loss under matched conditions has been obtained so far through careful parameter setting of tunable receivers, notably the mentioned Kalson's and KWA detectors; here we propose a different approach which, remarkably,  can cover levels of robustness that are not possible with such state-of-the-art receivers.
The idea in \eqref{eq:test} is in fact to add to the $H_1$ hypothesis a signal $\bm{\theta}$ that makes $H_1$ more plausible, hence hopefully the detector more robust to mismatches on the nominal steering vector $\bv$. To this aim, a design matrix $\bSigma$ is considered, multiplied by an unknown factor  $\nu$; in doing so, in case of matched signature no component will be  likely found along $\bSigma$ and the conventional case is recovered, i.e., the estimated value for $\nu$ will be likely zero; conversely, if the mismatch causes some leakage of the signal that is captured by $\bSigma$, the detector would tend to decide for $H_1$ more likely.

 If we suppose that $\bm{\theta}$ is independent of $\bn$,
the resulting hypothesis testing problem turns out to be
\begin{equation}
\left\{
\begin{array}{ll}
H_{0}: & \bz  \thicksim {\cal CN}_N (\bzero, \bC) \\ 
H_{1}: & \bz \thicksim {\cal CN}_N (\alpha \bv, \bC+ \nu \bm{\Sigma}).
\end{array}
\right. 
\label{modifiedtest}
\end{equation}

In summary, the contribution of this paper is threefold:
\begin{itemize}
\item A new hypothesis test \eqref{modifiedtest}  for radar detection is introduced, which is a possible generalization of  \eqref{eq:test1ord} and  
\eqref{eq:test2ord}.
\item The (one-step) GLRT is derived for two choices of the design matrix 
$\bm{\Sigma}$. A third parametric detector is obtained by an ad-hoc generalization of one of such GLRTs. Two-step GLRT-based detectors have been presented in our preliminary work \cite{CISS2018}. The detectors derived here have different complexity and behavior.
\item A thorough analysis of the proposed detectors is performed, also deriving closed-form expressions for the $P_{fa}$  for two of the detectors and showing that they have the desirable constant false alarm rate (CFAR) property. The performance assessment reveals that 
the proposed approach can guarantee negligible loss under matched conditions with respect to Kelly's detector while providing diversified degrees of robustness depending on chosen parameters.
\end{itemize}

The paper is organized as follows:  Section~II is devoted to the derivation of the GLRTs (with some proofs and lengthy manipulations in the Appendices). Section~III addresses the analysis of the detectors. We conclude in Section~IV.

\section{GLRTs for point-like targets}\label{sec:GLRTs}

In this section, we derive robust detectors employing the GLRT. To this end,
we consider the following binary hypothesis testing problem
$$
\left\{
\begin{array}{lll}
H_0:  & \bz \sim {\cal CN}_N(\bzero, \bC) & \\
 &  \bor_k \sim {\cal CN}_N(\bzero, \bC), & k=1, \ldots, K  \\ \\
H_1:  & \bz \sim {\cal CN}_N(\alpha  \bv, \nu \bSigma+ \bC) & \\
& \bor_k \sim {\cal CN}_N(\bzero, \bC), & k=1, \ldots, K
\end{array}
\right.
$$
where we recall that
the training data $\bor_k$
form a set of  $K \geq N$ independent (and identically distributed) vectors\footnote{The condition $K \geq N$ ensures that the sample covariance matrix based on training data has full rank with probability one \cite{Kelly}.}, independent also of $\bz$, free of target echoes, and  sharing with the CUT the statistical characteristics of the noise.
The positive definite matrix $\bC$, $\nu \geq 0$, and $\alpha \in \C$
are unknown quantities while $\bv \in \C^{N \times 1}$ is a known vector.
As to the (Hermitian) positive semidefinite matrix $\bSigma$, it might be either known or unknown.  In the former case,  $\bSigma$ reflects our knowledge about the random variations of the target steering vector around its nominal value $\bv$.  On the contrary, an unknown $\bSigma$ implies that no specific knowledge is available. In such a case, obviously $\bSigma$ cannot be estimated from a single snapshot (data from the CUT) and therefore one needs to make some further assumptions. The latter are mostly dictated by pragmatism so that the detection problem remains identifiable and mathematically tractable.  Indeed, $\bSigma$ should be viewed as a means to model the structure of the variations around $\bv$ while $\nu$ captures their amplitudes. 
 In any case, one has to look at the ultimate performance for judging the goodness of each choice. In the sequel, we will assume that either $\bSigma$ is a known rank-one matrix or 
$\bSigma=\bC$ (so that, although unknown, it does not introduce  new unknowns).

\subsection{Case 1: derivation of the GLRT for $\bSigma=\bu \bu^{\dagger}$ \label{GLRT-uuH} }\label{sec:case2}

In this section, we address the case that $\bSigma$ is a rank-one matrix.
First observe that {\em the case
$\bm{\Sigma} = \bv\bv^{\dagger}$,  which enforces the target signature  in both the mean and covariance\footnote{This amounts to assuming that $\bz = (\alpha + \epsilon) \bv + \bn$ where $\epsilon$ is a zero-mean complex normal random variable with unknown power $\nu$, independent of $\bn$. 
},
returns Kelly's detector}, as discussed later in this section.
We are instead more interested in a vector  $\bu$ aimed at making the $H_1$ hypothesis more plausible under mismatches. 
 Intuitively, a design matrix $\bm{\Sigma} = \bv_m\bv_m^{\dagger}$,  where $\bv_m$ is a slightly mismatched steering vector, is  a possible way to introduce some robustness without appreciable loss under matched conditions; nonetheless, other choices are possible, hence we provide a derivation for generic $\bSigma=\bu \bu^{\dagger}$.

The corresponding GLRT  is given by
\be
\Lambda(\bz, \bS)=
\frac{
\displaystyle{\max_{\bC >0} \max_{\nu \geq 0} \max_{\alpha \in \C} f_1( \bz, \bS | \bC, \nu, 
\alpha)
}}{
\displaystyle{\max_{\bC >0} f_0( \bz, \bS | \bC)
}}
\test \eta
\label{eq:1S-GLRT-u}
\ee
where 
\begin{equation}
f_1( \bz, \bS | \bC, \nu, \alpha) =
\frac{e^{- \mbox{tr} \left[ \left(\nu \bu \bu^\dag + \bC \right)^{-1}
\left( \bz - \alpha \bv \right) \left( \bz - \alpha \bv \right)^{\dagger} 
+ \bC^{-1} \bS \right]
}}{\pi^{N(K+1)} \det^{K}(\bC)\det(\nu \bu \bu^{\dagger} + \bC)}  
\label{eq:pdf_case_2_under_H1}
\end{equation}
and
$$
f_0( \bz, \bS | \bC) =
\frac{1}{\pi^{N(K+1)}} \frac{1}{\det^{K+1}(\bC)} 
e^{- \mbox{tr} \ \left\{ \bC^{-1} \left[  
 \bz \bz^{\dagger}
 + \bS \right] \right\} }
 $$
are the probability density functions (PDFs)
of 
$\bz, \bor_1, \ldots, \bor_{K}$
under $H_1$ and $H_0$, respectively,
and
$\bS/K$ the sample covariance matrix computed on the secondary data, i.e.,
$$
\bS=\sum_{k=1}^{K}  
\bor_k \bor_k^{\dagger}.
\label{eq:S}
$$
As to $\eta$, it is the detection threshold to be set according to the desired probability of false alarm ($P_{fa}$).
Maximization over $\bC$ of the likelihood under $H_0$
can be performed as in \cite{Kelly};
in fact, the maximizer of the likelihood is given by
$$
\widehat{\bC}_0
=
\frac{1}{K+1} \left[ \bz \bz^{\dagger} + \bS \right].
$$
It follows that
\be
f_0( \bz, \bS | \widehat{\bC}_0) =
\left( \frac{K+1}{e \pi} \right)^{N(K+1)} 
\left[ \det
\left( \bz \bz^{\dagger} + \bS \right)
\right]^{-(K+1)}.
\label{eq:pdf_under_H0}
\ee

Computation of the numerator of the GLRT
and,
in particular, maximization of $f_1( \bz, \bS | \bC, \nu, \alpha)$ with respect to $\bC$ is less standard. However, it can be conducted exploiting the following result (a special case of the one  in \cite{Besson_collaboration}).
\begin{proposition}
Assume that $\bu$ is a unit-norm vector (and that $K \geq N$).
Then, for the likelihood under the $H_1$ hypothesis, given by equation~(\ref{eq:pdf_case_2_under_H1}), 
the following equality holds true
\begin{align*}
l_{\max} &= \max_{\bC >0, \nu \geq 0, \alpha \in \C} f_1( \bz, \bS | \bC, \nu, \alpha) 
\\ &=
\max_{b \geq 0} \max_{\alpha \in \C} 
\frac{\left( \frac{K+1}{\pi e} \right)^{N(K+1)} (1+b)^{-1}}{\left[\det(\bS) \! \left( 1\! +\! \left( \bz \!-\! \alpha \bv \right)^{\dagger} \bS^{-1} \! \left( \bz \!-\! \alpha \bv \right) \right)\right]^{K+1}}
\\ &\times
\left[ \frac{\bu^{\dagger} \left(\bS + (1+b)^{-1} \left( \bz - \alpha \bv \right)\left( \bz \!-\! \alpha \bv \right)^{\dagger} \right)^{-1}\!\! \bu
}{
\bu^{\dagger} \left(\bS + \left( \bz - \alpha \bv \right)\left( \bz - \alpha \bv \right)^{\dagger} \right)^{-1} \!\! \bu
}
\right]^{\!\! K+1}
\end{align*}
where $b = \nu \bu^{\dagger} \bC^{-1} \bu$.
\end{proposition}
\medskip

\noindent
\textbf{Proof} 
The result is a special case of the one in \cite{Besson_collaboration} letting therein, in equation~(39), $\bR=\bC$,
$T_p=1$, $T_s=K$ (hence, $T_t=K+1$), $M=N$, $\bX=\bz - \alpha \bv$, $\bS_y=\bS$, $\bv=\bu$, $P=\nu$.
\phantom{x} \hfill \textbf{Q.E.D.}
\bigskip

In the following, we will denote by $l(b,\alpha)$ the partially-compressed likelihood under $H_1$, i.e.,
\begin{align*}
l(b,\alpha)
&=
\frac{\left( \frac{K+1}{\pi e} \right)^{N(K+1)}(1+b)^{-1}}{\det^{K+1}(\bS) \left( 1+ \left( \bz - \alpha \bv \right)^{\dagger} \bS^{-1} \left( \bz - \alpha \bv \right) \right)^{K+1}} 
\\ &\times
\left[ \frac{\bu^{\dagger} \!\! \left(\bS + (1+b)^{-1} \! \left( \bz - \alpha \bv \right)\left( \bz - \alpha \bv \right)^{\dagger} \right)^{-1} \!\! \bu
}{
\bu^{\dagger} \left(\bS + \left( \bz - \alpha \bv \right)\left( \bz - \alpha \bv \right)^{\dagger} \right)^{-1} \bu
}
\right]^{K+1} \!\!\!\!.
\end{align*}
As shown in Appendix~A, it can also be re-written as
\begin{align}
\nonumber
l(b,\alpha) &=
\frac{\left( \frac{K+1}{\pi e} \right)^{N(K+1)} (1+b)^{-1}}{\det^{K+1}(\bS) \left( 1+b+ 
\left\| \tilde{\bz} - \alpha \tilde{\bv} \right\|^2 
\right)^{K+1}} 
\\ &\times
\left[ 
\frac{
(1+b) +
\left\| \Putildeperp \left( \tilde{\bz} - \alpha \tilde{\bv} \right) \right\|^2
}{
1
+ 
\left\| \Putildeperp \left( \tilde{\bz} - \alpha \tilde{\bv} \right) \right\|^2
}
\right]^{K+1}
\label{eq:lprime}
\end{align}
where $\tilde{\bz}=\bS^{-1/2} \bz$, $\tilde{\bv}=\bS^{-1/2} \bv$, and $\tilde{\bu}=\bS^{-1/2} \bu$
are the ``whitened'' versions of $\bz$, $\bv$, and $\bu$, respectively, and we recall that $\Putildeperp$ is the (orthogonal) projection matrix onto the orthogonal complement of the subspace spanned by  $\tilde{\bu}$.

For $\bu=\bv$, $l(b,\alpha)$ can be easily maximized and the GLRT
is Kelly's detector (ref. Appendix A).
If, instead, $\bu \neq \bv$, maximization of the 
above partially-compressed likelihood with respect to $\alpha \in \C$ can be restricted to a proper 
``segment'' of the complex plane.
In fact, the following result holds true.
\begin{proposition}\label{theorem_caratterizza_minimizer}
Let $\Putildeperp \tilde{\bv} \neq \bzero$. The maximum of the function $l(b,\alpha)$ with respect to $\alpha \in \C$, given $b \geq 0$, 
is attained over the segment whose endpoints are
$\alpha_1$ and $\alpha_2$,
defined as
$$
\alpha_1= \arg \min_{\alpha} \left( \tilde{\bz} - \alpha \tilde{\bv} \right)^{\dagger} \left( \tilde{\bz} - \alpha 
\tilde{\bv} \right) = \left( \tilde{\bv}^{\dagger} \tilde{\bv} \right)^{-1} \tilde{\bv}^{\dagger} \tilde{\bz}
\label{eq:alpha_1}
$$
$$
\alpha_2= \arg \min_{\alpha} \left\| \Putildeperp \left( \tilde{\bz} - \alpha \tilde{\bv} \right) \right\|^2 =
\left( \tilde{\bv}^{\dagger} \Putildeperp \tilde{\bv} \right)^{-1} \tilde{\bv}^{\dagger} \Putildeperp \tilde{\bz}.
\label{eq:alpha_2}
$$
\end{proposition}
\medskip

\noindent
\textbf{Proof} 
See Appendix~A.
\bigskip

From the proposition above, $\alpha$ can be re-written as
$$
\alpha(t)=t \alpha_1+\left(1-t\right)\alpha_2, \qquad t \in \left[0,1\right].
$$
Accordingly, we have 
$$
|\alpha(t)-\alpha_1|^2=|t \alpha_1+\left(1-t\right)\alpha_2 -\alpha_1|^2=
\left(1-t\right)^2 |\alpha_2 - \alpha_1|^2,
$$
$$
|\alpha(t)-\alpha_2|^2=|t \alpha_1+\left(1-t\right)\alpha_2 -\alpha_2|^2=
t^2 |\alpha_1 - \alpha_2|^2;
$$
then, using equations~(\ref{eq:lvsb_and_alpha}), (\ref{eq:norma_alpha}),
and (\ref{eq:norma_pro_alpha}), together with the above two equations,
yields
\begin{align*}
l(b,\alpha(t)) \! &= \left( \frac{K+1}{\pi e} \right)^{N(K+1)}
\frac{1}{\det^{K+1}(\bS)}
\\ &\times
l_1\left(b, \left\| \tilde{\bz} - \alpha(t) \tilde{\bv} \right\|^2\right) 
l_2\left(b, \left\| \Putildeperp \left( \tilde{\bz} - \alpha(t) \tilde{\bv} \right) \right\|^2\right)
\\ &=
\left( \frac{K+1}{\pi e} \right)^{N(K+1)}
\frac{1}{\det^{K+1}(\bS)}
\\ &\times
l_1\! \left(b, \tilde{\bz}^{\dagger}\Pvtildeperp \tilde{\bz} + \left(1-t\right)^2 |\alpha_2 - \alpha_1|^2 \tilde{\bv}^{\dagger} \tilde{\bv}\right) 
\\ &\times
l_2\! \left(b, \tilde{\bz}^{\dagger}  \Putildeperp \Pcomplex \Putildeperp \tilde{\bz}
+ t^2 |\alpha_1 - \alpha_2|^2 \tilde{\bv}^{\dagger} \Putildeperp \tilde{\bv}\right)
\end{align*}
with $l_1(\cdot,\cdot)$
and $l_2(\cdot,\cdot)$
given in equation~(\ref{eq:def_l12}). 
Summarizing, the GLRT can be written as
\[
\frac{\max_{b \geq 0} \max_{t \in [0,1]} l(b,\alpha(t))
}{
\left( \frac{K+1}{\pi e} \right)^{N(K+1)} 
\displaystyle{ \frac{1}{\det^{K+1} \left( \bS \right)} }
\left( 1+  \bz^{\dagger} \bS^{-1} \bz \right)^{-(K+1)}
}
\test \eta
\]
and also as
\begin{align}
&\max_{b \geq 0} \max_{t \in [0,1]}
\frac{
l_1\left(b, \tilde{\bz}^{\dagger}\Pvtildeperp \tilde{\bz} + \left(1-t\right)^2 R^2 \tilde{\bv}^{\dagger} \tilde{\bv}\right) 
}{\left( 1+  \bz^{\dagger} \bS^{-1} \bz \right)^{-(K+1)}} \nonumber
\\ &\times
l_2\left(b, \tilde{\bz}^{\dagger}  \Putildeperp \Pcomplex \Putildeperp \tilde{\bz}
+ t^2 R^2 \tilde{\bv}^{\dagger} \Putildeperp \tilde{\bv}\right)
\test \eta
\label{eq:GLRT_Sigma=uu^H}
\end{align}
where $R^2=|\alpha_1 - \alpha_2|^2$.
Equation \eqref{eq:GLRT_Sigma=uu^H} is a computationally more convenient alternative for the GLRT in the case $\bSigma=\bu \bu^{\dagger}$, with $\bu$ any chosen vector. This rewriting is in fact less demanding than using $l(b,\alpha)$ for the numerator (partially-compressed likelihood under $H_1$): in particular, the maximization over $\alpha\in\C$ of $l(b,\alpha)$ (e.g., using equation \eqref{eq:lprime}) would require a complex-valued (two-dimensional) unbounded optimization instead of the simple scalar search of $t$ in the finite interval $[0,1]$ appearing in \eqref{eq:GLRT_Sigma=uu^H}.
Moreover, it is reasonable to constraint $b$ to belong to a finite interval, say $[0,b_{\max}]$.
This further reduces the ultimate complexity and makes it possible to implement the detector using a numerical algorithm with box constraints or, alternatively, a grid search.

\subsection{Case 2: derivation of the GLRT for $\bSigma=\bC$}

\noindent
In this case $\bSigma$ is unknown, but equal to the covariance matrix of the noise $\bC$. Clearly, since the latter is unrelated to the sources of error producing the steering vector mismatch that we hope to capture through the random term $\bm{\theta}$, a direct physical meaning for $\bSigma=\bC$ is missing; however, such a choice is mathematically convenient and leads to a detector with sound properties, as shown later.
Also,  if $\bC$ were known, the GLRT would process 
the whitened received signal and the choice
$\bm{\Sigma} = \bC$ would be tantamount to modeling the random component to be added to the (whitened) steering vector as a white term. This is reasonable when no a-priori knowledge on the nature of the mismatch is available, hence $\bSigma=\bC$ is actually a conservative (less informative) choice.\footnote{Please also recall the general rationale of test \eqref{eq:test} and the discussion at the beginning of Section \ref{sec:GLRTs}.} 

The corresponding GLRT is
\be
\Lambda(\bz, \bS)=
\frac{
\displaystyle{\max_{\bC >0} \max_{\nu \geq 0} \max_{\alpha \in \C} f_1( \bz, \bS | \bC, \nu, 
\alpha)
}}{
\displaystyle{\max_{\bC >0} f_0( \bz, \bS | \bC)
}}
\test \eta
\label{eq:1S-GLRT-C}
\ee
where 
\begin{equation*}
f_1( \bz, \bS | \bC, \nu, \alpha) =
\frac{e^{- \mbox{tr} \ \left\{ \bC^{-1} \left[  
\frac{1}{1+\nu}
\left( \bz - \alpha \bv \right)\left( \bz - \alpha \bv \right)^{\dagger}
 + \bS \right] \right\} }}{\pi^{N(K+1)}(1+\nu)^N \det^{K+1}(\bC)}
\end{equation*}
is the PDF of 
$\bz, \bor_1, \ldots, \bor_{K}$
under $H_1$
while
the denominator of \eqref{eq:1S-GLRT-C}
is given by \eqref{eq:pdf_under_H0}.
Again $\eta$ is the detection threshold for the desired $P_{fa}$.
We also recall that
$\bS$ is $K$ times the sample covariance matrix computed on the secondary data.

Maximization over $\bC$ of the likelihood under $H_1$
can be performed as in \cite{Kelly};
 in fact, we have that
$$
\widehat{\bC}_1(\nu, \alpha)
=
\frac{1}{K+1} \left[ \frac{1}{1+\nu}
\left( \bz - \alpha  \bv \right) \left( \bz - \alpha  \bv \right)^{\dagger} + \bS \right].
$$
Thus, 
the partially-compressed likelihood 
under $H_1$
becomes
\begin{eqnarray*}
&&f_1( \bz, \bS | \widehat{\bC}_1(\nu, \alpha), \nu, \alpha) =
\left( \frac{K+1}{e \pi} \right)^{N(K+1)} 
\\ &\times&\!\!\!\!\!\! \frac{1}{(1+\nu)^N}
\left[ \det\!
\left(\! \frac{1}{1+\nu}
\left( \bz - \alpha  \bv \right) \!\left( \bz - \alpha  \bv \right)^{\dagger} + \bS\! \right)
\right]^{-(K+1)}.
\end{eqnarray*}
Plugging the above expressions into equation~(\ref{eq:1S-GLRT-C}) yields
$$
\Lambda^{\frac{1}{K+1}}(\bz, \bS) 
=
\frac{
1+ \left\|  \tilde{\bz}  \right\|^2 
}{
\displaystyle\min_{\nu \geq 0, \alpha \in \C}
(1+\nu)^{\frac{N}{K+1}} \!\left[ 1+
\frac{1}{1+\nu} \left\|  \tilde{\bz} - \alpha  \tilde{\bv} \right\|^2
\right]
}
$$
where 
$\tilde{\bz} = \bS^{-1/2} \bz$, $\tilde{\bv}= \bS^{-1/2} \bv$.
Moreover, minimization over $\alpha$ leads to 
\cite{Horn-Johnson}
$$
\Lambda^{\frac{1}{K+1}}(\bz, \bS) =
\frac{
 1+ \|  \tilde{\bz}  \|^2
}{
\displaystyle\min_{\nu \geq 0} 
(1+\nu)^{\frac{N}{K+1}} \left[ 1+
\frac{1}{1+\nu} \left\| \Pvtildeperp \tilde{\bz} \right\|^2
\right]
}.
$$
Minimization of denominator with respect to $\nu$ can be easily accomplished 
using the following proposition.

\begin{table*}
\caption{Summary of proposed detectors for test \eqref{modifiedtest} and comparison with the classical approach (test \eqref{eq:test1ord}).}
\begin{center}
\begin{tabular}{|c|c|c|c|c|c|}
\hline
& no random term  & general $\bSigma$ & $\bSigma=\bu\bu^\dag$  & $\bSigma=\bv\bv^\dag$ & $\bSigma=\bC$ \\
& (classical test \eqref{eq:test1ord}) & & (general $\bu$) & & \\
\hline
 one-step GLRT & Kelly's detector & -- & eq. \eqref{eq:GLRT_Sigma=uu^H}  & Kelly's detector & eq. \eqref{eq:1S-GLRT-5C}; parametric generalization eq. \eqref{eq:parametric_detector} \\
 two-step GLRT  & AMF & eq. (9) in \cite{CISS2018} & eq. (9) in \cite{CISS2018} & AMF & eq. (11) in \cite{CISS2018} \\
\hline
\end{tabular}
\end{center}
\label{tabella}
\end{table*}%

\begin{proposition}
The function
$$
f(\nu)= (1+\nu)^{\frac{N}{K+1}} \left( 1 + \frac{a}{1+\nu} \right), \quad a >0, K \geq N
$$
admits a unique minimum over $[0, +\infty)$
at
$$
\widehat{\nu}= 
\left\{
\begin{array}{ll}
\left( \frac{K+1}{N} -1 \right) a-1, & \left( \frac{K+1}{N} -1 \right) a-1 > 0 \\
0, & \mbox{otherwise}
\end{array}
\right.
$$
given by
$
f_{\min} \!=\!
\left\{
\begin{array}{ll}
\left( \frac{K+1-N}{N} a \right)^{\!\frac{N}{K+1}} \!\!
\frac{K+1}{K+1-N}, & \frac{K+1-N}{N} a > 1  \\
1+a, & \mbox{otherwise}
\end{array}
\right.
$.
\label{eq:GLRT-A}
\end{proposition}
\medskip

\noindent
\textbf{Proof} 
{See Appendix B.}
\bigskip

Thus, it follows that the GLRT can be written as
\be
\Lambda_0(\bz, \bS) = \Lambda^{\frac{1}{K+1}}(\bz, \bS) 
\test \eta
\label{eq:1S-GLRT-5C}
\ee
where
$$
\Lambda_0(\bz, \bS) =
\left\{
\begin{array}{ll}
\frac{\left(1+\left\|  \tilde{\bz}  \right\|^2\right) \left( 1-\frac{1}{\zeta} \right)
}{\left[ \left(\zeta -1 \right)
\left\| \Pvtildeperp \tilde{\bz} \right\|^2 \right]^{\frac{1}{\zeta}} }, &
\left\| \Pvtildeperp \tilde{\bz} \right\|^2 > \frac{1}{\zeta-1}  \\
\frac{1+ \left\|  \tilde{\bz}  \right\|^2}{ 1+ \left\| \Pvtildeperp \tilde{\bz} \right\|^2}, & \mbox{otherwise}
\end{array}
\right.
$$
$\zeta=\frac{K+1}{N}$,
and $\eta$ denotes a modification of the original threshold.
The $P_{fa}$ of the test and the characterization of the decision statistic under $H_1$ can be obtained in closed form, and are reported in Appendix B. 
Remarkably, in the appendix it is shown that the detector has the desirable CFAR property.

Notice that, conditionally to
\be
\left\| \Pvtildeperp \tilde{\bz} \right\|^2 < \frac{1}{\zeta-1}
\label{eq:condition}
\ee
$\Lambda_0(\bz,\bS)$ is equivalent to Kelly's statistic; it turns out that the detector robustness comes from $\Lambda_0(\bz,\bS)$ under the condition complementary  to (\ref{eq:condition}).
Thus, it seems possible to promote robustness by decreasing the probability to select ``Kelly's statistic''. In particular, we propose to   replace $\zeta$ in (\ref{eq:condition}) with 
\begin{equation}
\zeta_{\epsilon}=\frac{K+1}{N}(1+\epsilon), \quad \epsilon \geq 0. \label{eq:zeta_eps}
\end{equation}
We made this ad-hoc substitution also on the decision statistic, obtaining the following parametric detector that thus generalizes \eqref{eq:1S-GLRT-5C} through \eqref{eq:zeta_eps}:
\be
\Lambda_\epsilon(\bz, \bS)  
\test \eta
\label{eq:parametric_detector}
\ee
where
$$
\Lambda_\epsilon(\bz, \bS) =
\left\{
\begin{array}{ll}
\frac{\left(1+\left\|  \tilde{\bz}  \right\|^2\right) \left( 1-\frac{1}{\zeta_{\epsilon}} \right)
}{\left[ \left(\zeta_{\epsilon} -1 \right)
\left\| \Pvtildeperp \tilde{\bz} \right\|^2 \right]^{\frac{1}{\zeta_{\epsilon}}} }, &
\left\| \Pvtildeperp \tilde{\bz} \right\|^2 > \frac{1}{\zeta_{\epsilon}-1} \\
\frac{1+ \left\|  \tilde{\bz}  \right\|^2}{ 1+ \left\| \Pvtildeperp \tilde{\bz} \right\|^2}, & \mbox{otherwise}.
\end{array}
\right.
$$
Such parametric detector encompasses detector (\ref{eq:1S-GLRT-5C}) for $\epsilon=0$ and eventually $\zeta_{\epsilon}=\zeta$. Moreover, we will show that detector \eqref{eq:parametric_detector} can outperform
detector (\ref{eq:1S-GLRT-5C})
in terms of robustness, without incurring additional loss under matched conditions. Finally, a closed form for its $P_{fa}$ is given in equation \eqref{eq:PFA-proposed} of Appendix B, which allows one to set the threshold to guarantee the desired $P_{fa}$ for any choice of $\epsilon$; for convenience, the curves for some relevant values of $\epsilon$ are drawn in Fig. \ref{fig:pfa_vs_eta}, also revealing that the $P_{fa}$ does not significantly change with $\epsilon$.
In Appendix B  it is furthermore proven that the proposed detector \eqref{eq:parametric_detector} (hence also \eqref{eq:1S-GLRT-5C}) possesses the CFAR property, i.e., its $P_{fa}$ does not depend on any unknown parameter (including the unknown noise covariance matrix $\bC$); additionally, it is proven that its performance in terms of $P_d$ depends only on the $\mbox{SNR}$ (equation~(\ref{eqn:defSNR})) and the cosine squared $\cos^2 \theta$ (equation~(\ref{eqn:defcostheta})) between the nominal steering vector and the actual one, which are the same performance parameters of Kelly's, AMF, Kalson's, and KWA detectors.  This remarkable fact corroborates the soundness and convenience of the choice $\bSigma=\bC$.
It also means that $P_d$ does not depend directly on noise-related parameters, namely the clutter power and correlation, but only through SNR and $\cos^2\theta$; as a consequence, the figures we will provide in Section \ref{sec:perf} ($P_d$ vs SNR for different $\cos^2\theta$) are quite representative of the general performance, despite they are obtained for a specific choice of $\bC$.

\begin{figure}
\centering
\includegraphics[width=8cm]{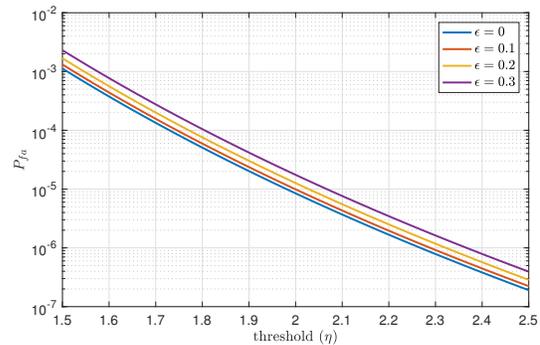}
\caption{Curves of $P_{fa}$ vs detection threshold, for different values of $\epsilon$.}\label{fig:pfa_vs_eta}
\end{figure}

\subsection{Discussion}

It is important to stress that the derivation of the GLRT for $\bSigma=\bC$ and, hence, the result is different from the detector proposed in \cite{ACE}.
More precisely, considering the derivation of the ACE as the GLRT to detect a possible coherent signal in presence of partially-homogeneous noise, given in \cite{ACE}, we highlight that
therein the parameter $\nu$ represents the possible mismatch
between the power of the noise in the data under test and that of secondary data. Herein, instead, $\nu$ enters the characterization of a possible random mismatch between
the actual and the nominal useful signal. For this reason, we have $1+\nu$ in place of $\nu$ under $H_1$
($\nu \bC$ stands for the covariance matrix of the random term)
and, in addition, $\nu$ is not present under the $H_0$ hypothesis. This difference in the formulation of the hypothesis testing problem makes our detector more robust than Kelly's detector \cite{Kelly} in presence of mismatched signals while it is well-known that the ACE is more selective than the latter (under the same operating conditions).

Table \ref{tabella} reports a summary of the proposed detectors for test \eqref{modifiedtest}, also in comparison with the classical approach (test \eqref{eq:test1ord}).\footnote{For completeness, we mention that \cite{CISS2018} provides the general derivation (for any $\bSigma$)  and a closed-form for $\bSigma=\bC$ following the two-step GLRT approach, namely deriving the GLRT for known $\bC$ and then using as replacement an estimate $\hat{\bC}$ (from secondary data); such detectors are however less powerful than one-step GLRTs.
In Appendix~C  we show that {\em for $\bSigma=\bv \bv^{\dagger}$ the two-step GLRT is equivalent to the AMF, whereas the proposed one-step GLRT for the same choice of $\bSigma$ is equivalent to Kelly's detector} (ref. Section~\ref{sec:case2}).}

 \begin{figure}
\centering
\subfigure[$K=24$]{
\includegraphics[width=7.5cm]{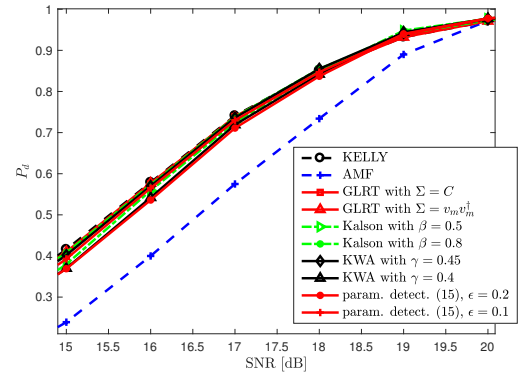}
}
\subfigure[$K=32$]{
\includegraphics[width=7.5cm]{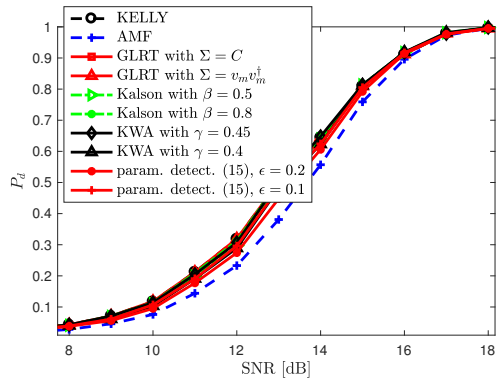}
}
\subfigure[$K=40$]{
\includegraphics[width=7.5cm]{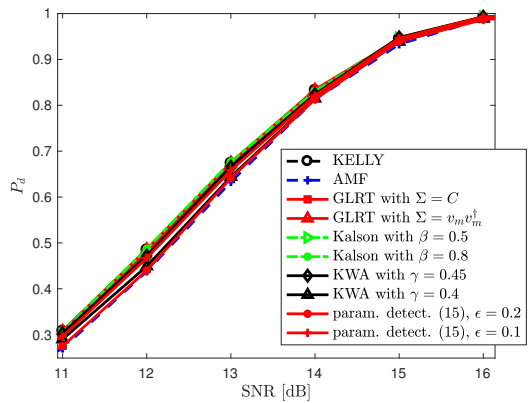}
}
\caption{$P_d$ vs SNR under matched conditions, for $P_{fa}=10^{-4}$.}\label{fig:0}
\end{figure}

\section{Performance analysis}\label{sec:perf}

\begin{figure}
\centering
\subfigure[$K=24$]{
\includegraphics[width=7.5cm]{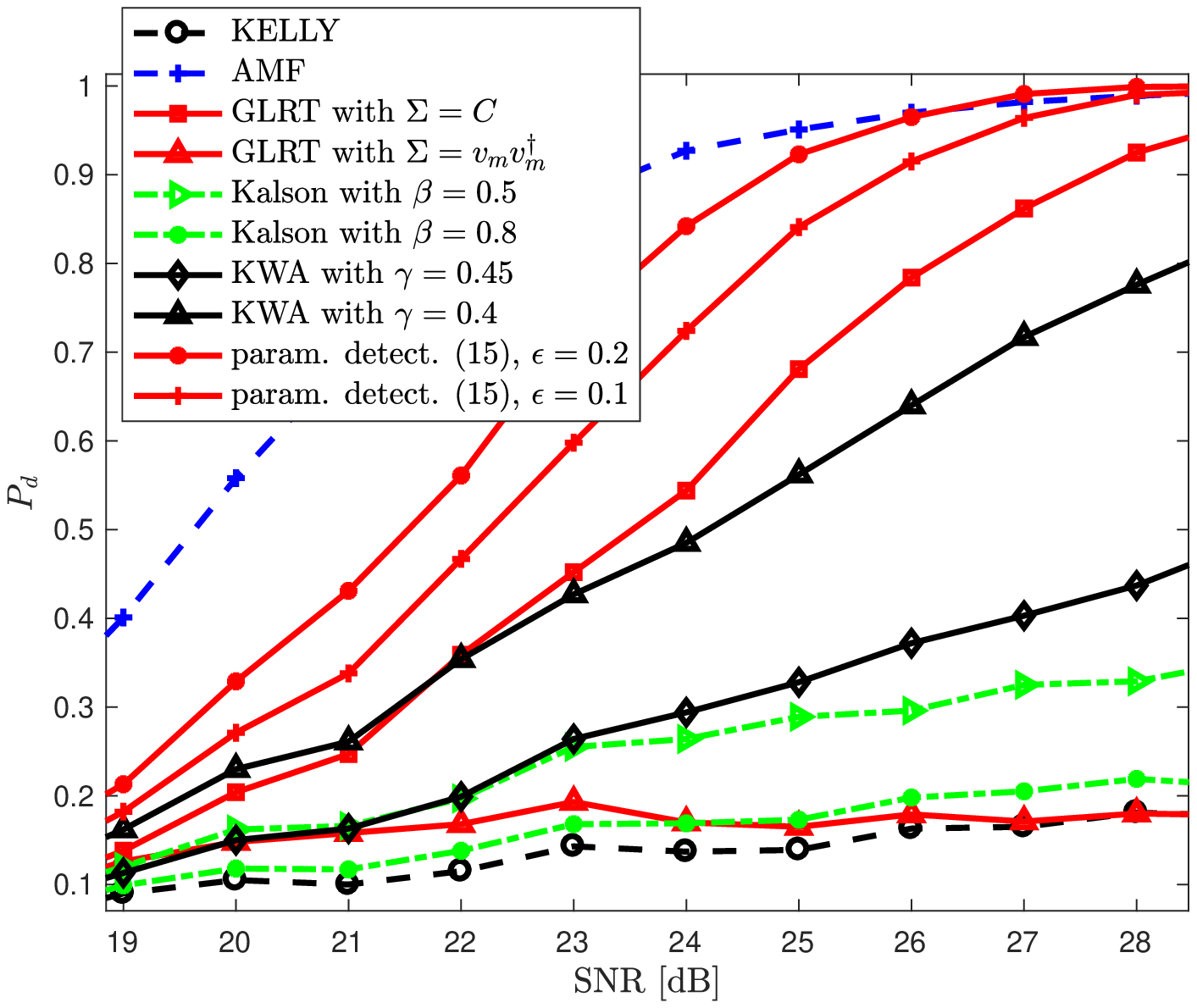}
}
\subfigure[$K=32$]{
\includegraphics[width=7.5cm]{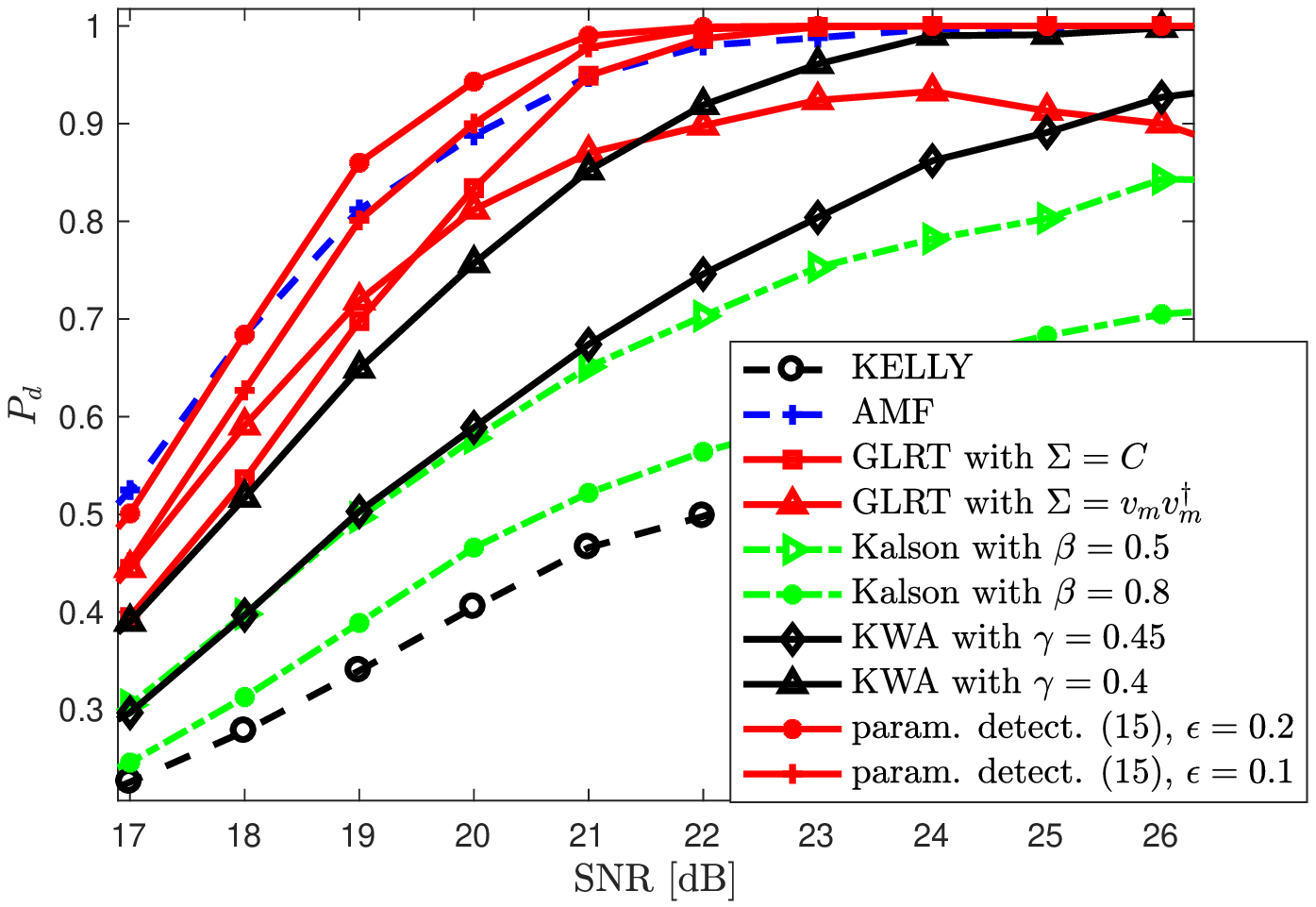}
}
\subfigure[$K=40$]{
\includegraphics[width=7.5cm]{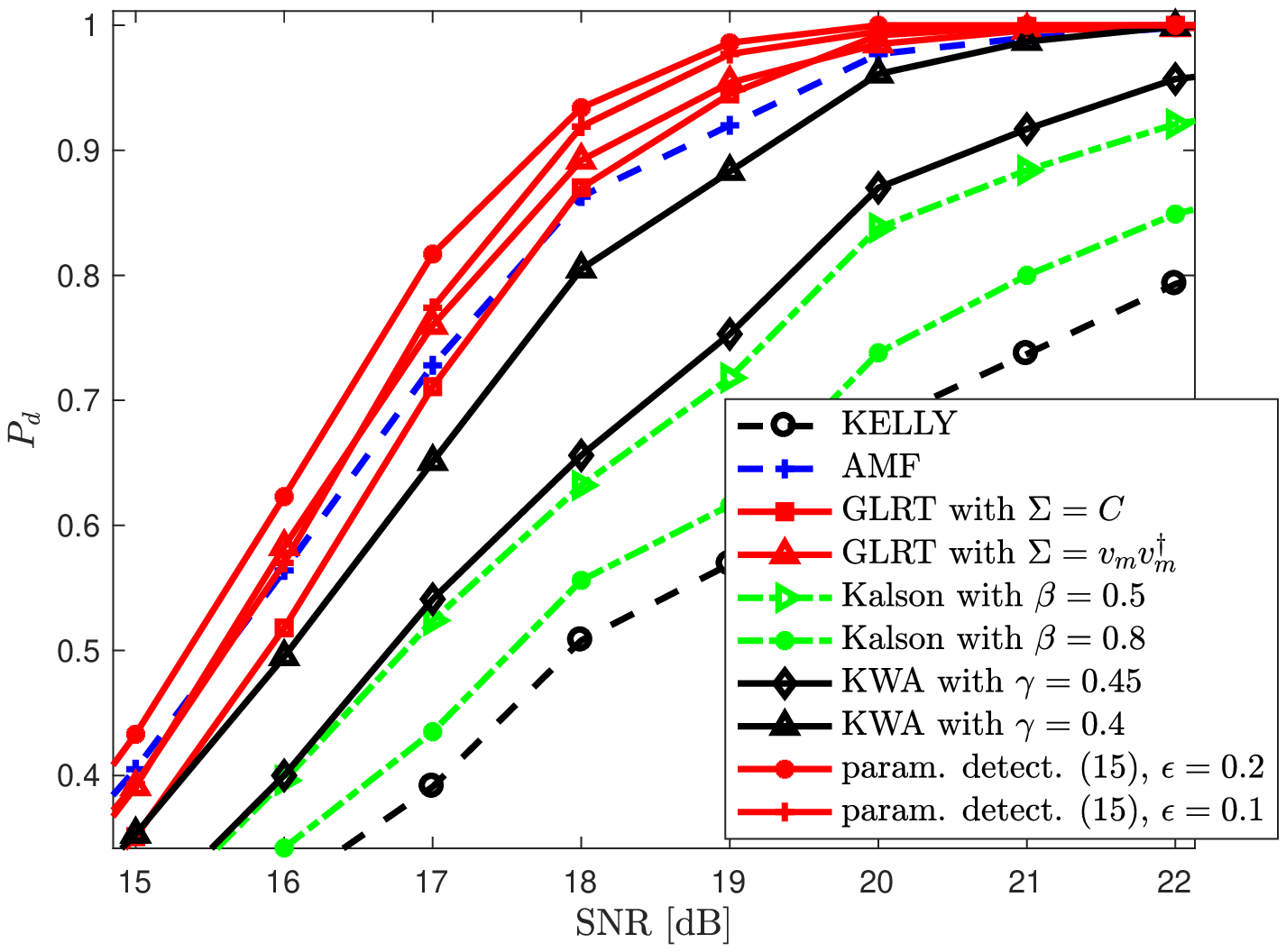}
}
\caption{$P_d$ vs SNR in case of mismatched steering vector, for $\cos^2 \theta \approx 0.46$ and  $P_{fa}=10^{-4}$.}\label{fig:4}
\end{figure}

 We assume a time steering vector, i.e., $\bv=[1\  e^{i2\pi f_d} \ \cdots \ e^{i2\pi (N-1)f_d} ]^{T}$, with $N=16$ and normalized Doppler frequency $f_d=0.08$, a small value such that the target competes with low pass clutter. 
The target amplitude  $\alpha$ is generated deterministically according to the SNR defined in equation \eqref{eqn:defSNR}.
To assess the robustness of the proposed detectors, we simulate a target with a mismatched steering vector, say $\bp$, having normalized Doppler frequency $f_d+\delta_f$ with $\delta_f \in\{0.2/N,0.4/N\}$. The values of $\delta_f$ will affect the values of the cosine squared of the angle between the nominal steering vector and the mismatched one, denoted by $\cos^2 \theta$ and given in equation~(\ref{eqn:defcostheta}), which is a standard way to quantify the mismatch in radar detection. Under matched conditions, of course, $\bp=\bv$ and $\cos^2 \theta=0$.

We model the noise component of $\bz$ and the $\bor_k$s as (independent) random vectors ruled by  a zero-mean, complex Gaussian distribution. As to the covariance matrix, we adopt as $\bC$ the sum of  a Gaussian-shaped clutter covariance matrix and white (thermal) noise 10 dB weaker, i.e.,
$\bC = \bR_c + \sigma_n^2 \bI_N$ with  the $(m_1,m_2)$th element of the matrix $\bR_c$ given by
$[\bR_c]_{m_1,m_2} \propto \exp\{- 2\pi^2\sigma_f^2(m_1-m_2)^2\}$ and $\sigma_f \approx 0.073$ (corresponding to a one-lag correlation coefficient of the clutter component equal to $0.9$).

In the following we assess the performance of the proposed  (one-step) GLRTs, considering both $\bSigma=\bv_m \bv_m^{\dagger}$ and 
$\bSigma=\bC$, and the parametric detector \eqref{eq:parametric_detector}.
The former detector assumes
as $\bv_m$  a slightly
mismatched version of $\bv$, in particular of $0.03/N$, and for the optimization with respect to $b$ we adopt a logarithmic search in the span $[0,b_{\max}=10^3]$; as regards the parametric detector \eqref{eq:parametric_detector}, we consider $\epsilon=0.1$ and $\epsilon=0.2$. It will be apparent from the simulations below that such choices guarantee negligible loss with respect to Kelly's detector under matched conditions.

A comparison is performed against the natural competitors for the problem at hand, that is the Kalson's \cite{Kalson} and KWA \cite{KWA} tunable receivers. Notice that for both such competitors, the level of robustness increases as the value of their tunable parameter decreases. For a fair comparison, only values that ensure practically the same performance of Kelly's detector under matched conditions are considered, in particular $\beta=0.5$ and $\beta=0.8$ for Kalson's detector and $\gamma=0.45$ and $\gamma=0.4$ for the KWA.
For reference, the AMF \cite{Kelly-Nitzberg},
is also reported in all figures, because it is a well-known robust detector; however, it is worth remarking that it is not always a competitor, since it generally experiences a performance loss under matched conditions with respect to Kelly's detector.

\begin{figure}
\centering
\subfigure[$K=24$]{
\includegraphics[width=7.5cm]{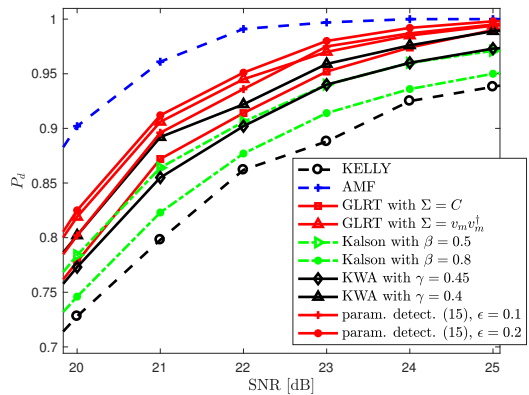}
}
\subfigure[$K=32$]{
\includegraphics[width=7.5cm]{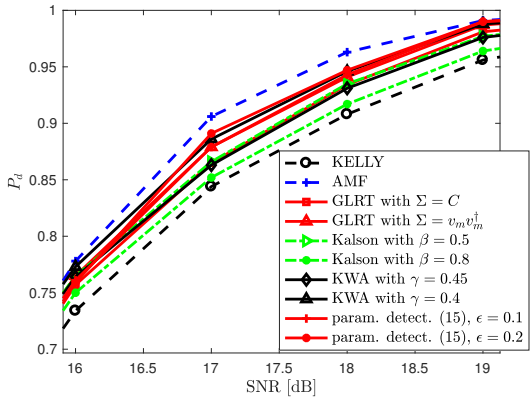}
}
\subfigure[$K=40$]{
\includegraphics[width=7.5cm]{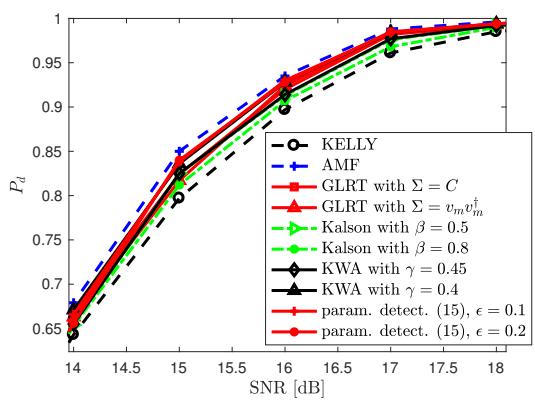}
}
\caption{$P_d$ vs SNR in case of mismatched steering vector, for $\cos^2 \theta \approx 0.83$ and  $P_{fa}=10^{-4}$.}\label{fig:2}
\end{figure}

A first set of simulations (Figs. \ref{fig:0}--\ref{fig:2}) refers to $P_{fa}=10^{-4}$. We assess performance by
Monte Carlo simulation with $100/P_{fa}$ independent trials to set the thresholds
and $10^3$ independent trials to compute the $P_ d$s (the probabilities to decide for $H_1$ when a useful signal is present).
Results under matched conditions are reported in Fig. \ref{fig:0} for different values of the number of secondary data, namely $K=24,32,40$ respectively (i.e., besides the typical  $K=2N$, also $1.5N$ and $2.5N$ are considered). From such figures, it is apparent that the proposed detectors and the tunable competitors (Kalson's and KWA) have essentially the same $P_d$ of Kelly's detector, while as known the AMF 
exhibits a performance loss, which becomes negligible only when the number of secondary data is large enough. Notice that the KWA for $\gamma=0.4$ and the detector \eqref{eq:parametric_detector} for $\epsilon=0.2$ start to experience some loss, especially for moderate SNR, hence  smaller values of their respective parameters are not considered for a fair comparison.

Fig. \ref{fig:4} reports the results under mismatch, in particular considering a cosine squared of the angle between the nominal steering vector and the mismatched one (given by equation~(\ref{eqn:defcostheta})) equal to $\cos^2 \theta \approx 0.46$ (obtained for  $\delta_f=0.4/N=0.025$).
It emerges that the proposed approach yields a family of robust receivers, whose performance in terms of robustness to mismatches depends on the design matrix $\bm{\Sigma}$ and number of secondary data $K$.
The behavior of the GLRT for $\bm{\Sigma} = \bC$
is very interesting, because it is (equivalent to Kelly's detector under matched conditions and) more robust than Kalson's and KWA receivers; the parametric detector   \eqref{eq:parametric_detector} with both settings of $\epsilon$
is more robust
than the GLRT
for $\bm{\Sigma} = \bC$ and can be even more robust than AMF, in particular for $K=32$, without the loss experienced by the latter under matched conditions. This ``win-win'' behavior is thus remarkably unique and not achievable with state-of-the-art receivers.
Notice that  the AMF can be considered a competitor  only for $K=40$ (where its performance loss under matched conditions compared to Kelly's detector becomes very small), but Fig. \ref{fig:4}(c) shows that the proposed detectors, in particular the GLRT for $\bSigma=\bC$ and the parametric detector \eqref{eq:parametric_detector} with both settings of $\epsilon$, are more robust than AMF; notice that this advantage comes  at no increase in computational complexity.

The GLRT with $\bSigma=\bv_m \bv_m^{\dagger}$ needs higher values of $K$ to achieve almost the same robustness of the other proposed receivers (Fig. \ref{fig:4}(c)); otherwise, it progressively rejects the $H_1$ hypothesis: for $K=32$ (Fig. \ref{fig:4}(b)) only for larger SNR (not shown in the figure), for $K=24$ (Fig.  \ref{fig:4}(a)) behaving as a selective receiver over the whole range of SNRs. As to Kalson's detector, for $\beta=0.8$ it is close to Kelly's detector (in particular for $K=24$).

For a reduced value of the mismatch, the differences between the proposed detectors and the competitors reduce. Fig.  \ref{fig:2} shows the mismatched case with $\cos^2 \theta \approx 0.83$ (obtained for  $\delta_f=0.2/N=0.0125$).
We observe that the parametric detector \eqref{eq:parametric_detector} still emerges as the most robust among the detectors that have negligible loss under matched conditions (compared to Kelly's detector) hence excluding AMF for $K=24,32$; for $K=40$, conversely, the  detector \eqref{eq:parametric_detector} and AMF are almost equivalent.

\begin{figure}
\centering
\includegraphics[width=7.5cm]{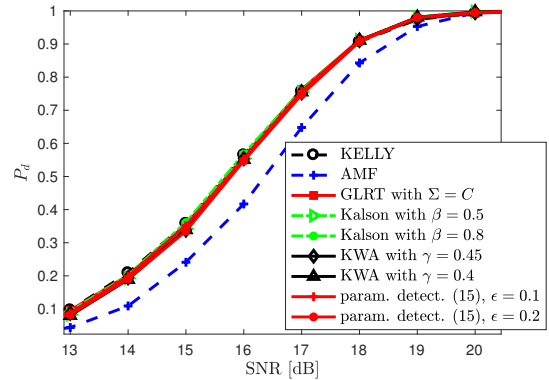}
\caption{$P_d$ vs SNR in case of matched steering vector, for $K=32$, and  $P_{fa}=10^{-6}$.}\label{fig:4_1e-6_rho09_match}
\end{figure}

\begin{figure}
\centering
\includegraphics[width=7.5cm]{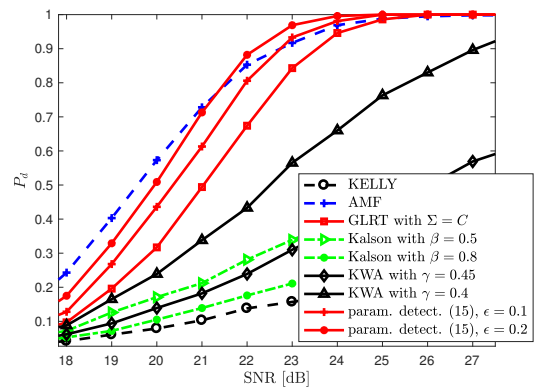}
\caption{$P_d$ vs SNR in case of mismatched steering vector, for $K=32$, $\cos^2 \theta \approx 0.46$, and  $P_{fa}=10^{-6}$.}\label{fig:4_1e-6_rho09_mismatch}
\end{figure}

A second set of simulations (Figs. \ref{fig:4_1e-6_rho09_match}--\ref{fig:4_1e-6_rho09_mismatch})
refers to $P_{fa}=10^{-6}$ and focuses on the behavior of the proposed GLRT for $\bm{\Sigma} = \bC$ and the parametric detector (\ref{eq:parametric_detector})  in comparison to Kalson's and KWA detectors; for reference, also the performance of Kelly's detector and AMF are shown. Notice that, for such a small $P_{fa}$, Monte Carlo simulations are prohibitively long, hence we can compare only detectors for which a closed-form expression is available for the relationship between
the threshold and the $P_{fa}$; for this reason the proposed GLRT with
$\bSigma=\bv_m \bv_m^{\dagger}$ cannot be included; besides, the detector with $\bm{\Sigma} = \bC$  and the parametric detector (\ref{eq:parametric_detector}) are simpler and show a more interesting behavior, hence are more appealing for applications.
Again we resort to Monte Carlo simulation and to $10^3$ independent trials to estimate the $P_d$s.
The values of the tunable parameters for the parametric detector and the competitors are chosen as before, because this guarantees  no appreciable loss under matched conditions as shown in Fig. \ref{fig:4_1e-6_rho09_match}. In Fig. \ref{fig:4_1e-6_rho09_mismatch} we report the results under mismatched conditions for $\cos^2 \theta \approx 0.46$ and $K=32$.
Comparing the result with the corresponding Fig.  \ref{fig:4}(b), the better performance of the proposed receivers are  even more evident.
We see that they are more robust than Kalson's and KWA detectors; in particular, the parametric detector with $\epsilon=0.2$ is also more robust than the AMF for large SNR values while Kalson's detector with $\beta=0.8$ approaches the selective behavior of Kelly's detector. 

\section{Conclusion}

We have proposed a novel family of robust radar receivers to detect the possible presence of a coherent return from a given cell under test mismatched with respect to the nominal steering vector. To this end, we have introduced a random component in addition to the deterministic useful signal; the random component has a given structure $\bSigma$, but its power $\nu$ is estimated from the data.  This likely produces an estimate of $\nu$ close to zero under matched conditions, thus achieving no loss compared to Kelly's detector (which is the benchmark for the considered problem).
We have solved the hypothesis testing problem resorting to the GLRT approach and assuming for  $\bSigma$ either a known rank-one matrix or the unknown covariance matrix of the noise. We have also introduced a parametric detector that naturally modifies the 
statistic of the latter GLRT and shares with it the CFAR property.
The performance analysis has shown that proposed detectors are a viable means to deal with mismatched signals;   in fact, compared to competitors that have negligible loss under matched conditions, the proposed approach guarantees an increased robustness.

Outgrowths of this work are the investigation of further possible structures for the design matrix $\bSigma$, and possibly the derivation of the one-step GLRT for the general case (which in fact is missing in Table \ref{tabella}). A different avenue for possible generalizations is robust detection of range-spread targets; to date, the GLRT for the proposed random perturbation approach has been derived only in the case $\bSigma=\bC$, which though has the same remarkable properties of the point-like detector derived here \cite{SP2020}. Finally, it would be interesting to ascertain whether receivers with the same detection power of Kelly's detector (under matched conditions) but controllable robustness or selectivity under mismatched conditions can be obtained through design approaches different from GLRT, namely inspired to machine learning tools.

\section*{Appendix A}
\renewcommand{\theequation}{A.\arabic{equation}}

\section*{Derivation of equation~(\ref{eq:lprime})}

First observe that, by the matrix inversion lemma, we have
\begin{align*}
& \bu^{\dagger} \left(\bS + (1+b)^{-1} \left( \bz - \alpha \bv \right)\left( \bz - \alpha \bv \right)^{\dagger} \right)^{-1} \bu
\\ &= \bu^{\dagger} \bS^{-1} \bu -   \frac{(1+b)^{-1} \left| \bu^{\dagger} \bS^{-1} \left( \bz - \alpha \bv \right) \right|^2 }{
1+ (1+b)^{-1} \left( \bz - \alpha \bv \right)^{\dagger} \bS^{-1} \left( \bz - \alpha \bv \right) }.
\end{align*}
Thus,  $l(b,\alpha)$ can be re-written as
\begin{align*}
& l(b,\alpha) =
\frac{\left( \frac{K+1}{\pi e} \right)^{N(K+1)} (1+b)^{-1}}{\det^{K+1}(\bS) \left( 1+ \left( \tilde{\bz} - \alpha \tilde{\bv} \right)^{\dagger} \left( \tilde{\bz} - \alpha 
\tilde{\bv} \right) \right)^{K+1}} 
\\ &\times
\left[ \frac{
\tilde{\bu}^{\dagger} \tilde{\bu} -  (1+b)^{-1}  \frac{ \left| \tilde{\bu}^{\dagger} \left( \tilde{\bz} - \alpha \tilde{\bv} \right) \right|^2 }{
1+ (1+b)^{-1}  \left( \tilde{\bz} - \alpha \tilde{\bv} \right)^{\dagger} \left( \tilde{\bz} - \alpha \tilde{\bv} \right) }
}{
\tilde{\bu}^{\dagger} \tilde{\bu} -  \frac{ \left| \tilde{\bu}^{\dagger} \left( \tilde{\bz} - \alpha \tilde{\bv} \right) \right|^2 }{
1+ \left( \tilde{\bz} - \alpha \tilde{\bv} \right)^{\dagger} \left( \tilde{\bz} - \alpha \tilde{\bv} \right) }
}
\right]^{K+1}
\end{align*}
where $\tilde{\bz}=\bS^{-1/2} \bz$, $\tilde{\bv}=\bS^{-1/2} \bv$, and $\tilde{\bu}=\bS^{-1/2} \bu$
are the whitened versions of $\bz$, $\bv$, and $\bu$, respectively.

Notice also that
\begin{align*}
&
\frac{
\tilde{\bu}^{\dagger} \tilde{\bu} -  (1+b)^{-1}  \frac{ \left| \tilde{\bu}^{\dagger} \left( \tilde{\bz} - \alpha \tilde{\bv} \right) \right|^2 }{
1+ (1+b)^{-1}  \left( \tilde{\bz} - \alpha \tilde{\bv} \right)^{\dagger} \left( \tilde{\bz} - \alpha \tilde{\bv} \right) }
}{
\tilde{\bu}^{\dagger} \tilde{\bu} -  \frac{ \left| \tilde{\bu}^{\dagger} \left( \tilde{\bz} - \alpha \tilde{\bv} \right) \right|^2 }{
1+ \left( \tilde{\bz} - \alpha \tilde{\bv} \right)^{\dagger} \left( \tilde{\bz} - \alpha \tilde{\bv} \right) }
}
\\ &=
\frac{
\tilde{\bu}^{\dagger} \tilde{\bu} -  \frac{ \left| \tilde{\bu}^{\dagger} \left( \tilde{\bz} - \alpha \tilde{\bv} \right) \right|^2 }{
1+ b+  \left( \tilde{\bz} - \alpha \tilde{\bv} \right)^{\dagger} \left( \tilde{\bz} - \alpha \tilde{\bv} \right) }
}{
\tilde{\bu}^{\dagger} \tilde{\bu} -  \frac{ \left| \tilde{\bu}^{\dagger} \left( \tilde{\bz} - \alpha \tilde{\bv} \right) \right|^2 }{
1+ \left( \tilde{\bz} - \alpha \tilde{\bv} \right)^{\dagger} \left( \tilde{\bz} - \alpha \tilde{\bv} \right) }
}
\\ &=
\frac{
(1+b) \tilde{\bu}^{\dagger} \tilde{\bu} +
\tilde{\bu}^{\dagger} \tilde{\bu}
\| \tilde{\bz} - \alpha \tilde{\bv}  \|^2-
\left| \tilde{\bu}^{\dagger} \left( \tilde{\bz} - \alpha \tilde{\bv} \right) \right|^2 
}{
\tilde{\bu}^{\dagger} \tilde{\bu} 
+ \tilde{\bu}^{\dagger} \tilde{\bu} 
\| \tilde{\bz} - \alpha \tilde{\bv} \|^2 -  \left| \tilde{\bu}^{\dagger} \left( \tilde{\bz} - \alpha \tilde{\bv} \right) \right|^2 
}
\\ &\times
\frac{ 1+ \left( \tilde{\bz} - \alpha \tilde{\bv} \right)^{\dagger} \left( \tilde{\bz} - \alpha \tilde{\bv} \right) 
}{
1+ b+  \left( \tilde{\bz} - \alpha \tilde{\bv} \right)^{\dagger} \left( \tilde{\bz} - \alpha \tilde{\bv} \right) 
}.
\end{align*}
It turns out that $l(b,\alpha)$ can be re-cast as 
\begin{align*}
&l(b,\alpha) =
\frac{\left( \frac{K+1}{\pi e} \right)^{N(K+1)}(1+b)^{-1}}{\det^{K+1}(\bS) \left( 1+b+ \left( \tilde{\bz} - \alpha \tilde{\bv} \right)^{\dagger} \left( \tilde{\bz} - \alpha 
\tilde{\bv} \right) \right)^{K+1}} 
\\ &\times
\left[ 
\frac{
(1+b) \| \tilde{\bu} \|^2 +
\| \tilde{\bu} \|^2
\| \tilde{\bz} - \alpha \tilde{\bv} \|^2 -
\left| \tilde{\bu}^{\dagger} \left( \tilde{\bz} - \alpha \tilde{\bv} \right) \right|^2 
}{
\| \tilde{\bu} \|^2
+ \| \tilde{\bu} \|^2 
\| \tilde{\bz} - \alpha \tilde{\bv} \|^2 -  \left| \tilde{\bu}^{\dagger} \left( \tilde{\bz} - \alpha \tilde{\bv} \right) \right|^2 
}
\right]^{K+1}
\end{align*}
and letting
$
\left| \tilde{\bu}^{\dagger} \left( \tilde{\bz} - \alpha \tilde{\bv} \right) \right|^2
=\| \tilde{\bu} \|^2 \left\| \Putilde \left( \tilde{\bz} - \alpha \tilde{\bv} \right) \right\|^2
$
it can be re-written in the form of equation~(\ref{eq:lprime}).

\section*{The GLRT with
$\bSigma=\bv\bv^\dag$ is the Kelly's detector}

If $\bu=\bv$ and, hence, $\Putildeperp \tilde{\bv}=0$, the partially-compressed likelihood under $H_1$ of equation (\ref{eq:lprime})
becomes
\begin{align*}
l(b,\alpha) &=
\frac{\left( \frac{K+1}{\pi e} \right)^{N(K+1)}(1+b)^{-1}}{\det^{K+1}(\bS) \left( 1+b+ 
\left( \tilde{\bz} - \alpha \tilde{\bv} \right)^{\dagger} \left( \tilde{\bz} - \alpha 
\tilde{\bv} \right)
\right)^{K+1}} 
\\ &\times
\left[ 
\frac{
(1+b) +
\left\| \Pvtildeperp \tilde{\bz} \right\|^2
}{
1
+ 
\left\| \Pvtildeperp \tilde{\bz} \right\|^2
}
\right]^{K+1}
\end{align*}
and
\begin{equation*}
\max_{\alpha} l(b,\alpha) =
\frac{(1+b)^{-1} \left( \frac{K+1}{\pi e} \right)^{N(K+1)}}{\det^{K+1}(\bS) \left( 1 + \left\| \Pvtildeperp \tilde{\bz} \right\|^2
\right)^{K+1}}.
\end{equation*}
As a consequence, we also have that
\begin{equation*}
\max_{b \geq0} \max_{\alpha} l(b,\alpha) =
\frac{\left( \frac{K+1}{\pi e} \right)^{N(K+1)}}{\det^{K+1}(\bS) \left( 1 + \left\| \Pvtildeperp \tilde{\bz} \right\|^2
\right)^{K+1}} 
\end{equation*}
and the GLRT, given by equation (\ref{eq:1S-GLRT-u}), can be written as
\[
\frac{1+ \bz^{\dagger} \bS^{-1} \bz }{
1 + \left\| \Pvtildeperp \tilde{\bz} \right\|^2
}
=
\frac{1+ \bz^{\dagger} \bS^{-1} \bz }{
1 + \bz^{\dagger} \bS^{-1} \bz
- \frac{\left| \bv^{\dagger} \bS^{-1} \bz \right|^2}{\bv^{\dagger} \bS^{-1} \bv}
}
\test \eta,
\]
with $\eta$ a proper modification of the original threshold, that is equivalent to Kelly's detector (we have also used equation (\ref{eq:pdf_under_H0}) for the compressed likelihood under $H_0$).

\section*{Proof of Proposition~\ref{theorem_caratterizza_minimizer}}

First observe that
$l(b,\alpha)$ 
can be written as
\begin{equation}
l(b,\alpha) =
\frac{l_1\!\left(b, \left\| \tilde{\bz} - \alpha \tilde{\bv} \right\|^2 \right) \,
l_2\!\left(b, \left\| \Putildeperp \left( \tilde{\bz} - \alpha \tilde{\bv} \right) \right\|^2\right)}{\left( \frac{K+1}{\pi e} \right)^{-N(K+1)} \det^{K+1}(\bS)}
  \label{eq:lvsb_and_alpha}
\end{equation}
with
\be
l_1(b, y_1) \!=\! \frac{(1+b)^{-1}}{
\left( 1+b+ y_1 \right)^{K+1}}, \,\,
l_2(b, y_2) \!=\! \left[ \frac{ (1+b) + y_2}{1+ y_2}\right]^{K+1}.
\label{eq:def_l12}
\ee
In particular, $l_1$ and $l_2$ are strictly decreasing function of 
$y_1$
and
$y_2$, respectively.
Moreover, $l_1$ attains its maximum at
$$
\alpha_1= \arg \min_{\alpha} \left( \tilde{\bz} - \alpha \tilde{\bv} \right)^{\dagger} \left( \tilde{\bz} - \alpha 
\tilde{\bv} \right) = \left( \tilde{\bv}^{\dagger} \tilde{\bv} \right)^{-1} \tilde{\bv}^{\dagger} \tilde{\bz}
$$
and
\begin{align}
\label{eq:norma_alpha}
\| \tilde{\bz} - \alpha \tilde{\bv} \|^2  &= \| \tilde{\bz} - \left( \alpha - \alpha_1 \right) \tilde{\bv} - \alpha_1 \tilde{\bv} \|^2
\\ \nonumber &=
\tilde{\bz}^{\dagger}\Pvtildeperp \tilde{\bz} + \left| \alpha - \alpha_1 \right|^2 \tilde{\bv}^{\dagger} \tilde{\bv}
\geq
\tilde{\bz}^{\dagger}\Pvtildeperp \tilde{\bz} 
\end{align}
where $\tilde{\bz} - \alpha_1 \tilde{\bv}= \Pvtildeperp \tilde{\bz}$.
Thus, $\forall x \geq \tilde{\bz}^{\dagger}\Pvtildeperp \tilde{\bz}$, the equation
$
\left( \tilde{\bz} - \alpha \tilde{\bv} \right)^{\dagger} \left( \tilde{\bz} - \alpha 
\tilde{\bv} \right) = x
$
is tantamount to 
$
\left| \alpha - \alpha_1 \right|^2 \tilde{\bv}^{\dagger} \tilde{\bv}=
x-\tilde{\bz}^{\dagger}\Pvtildeperp \tilde{\bz}
$
and is a circle centered at $\alpha_1$ of proper radius.
Similarly, $l_2$ attains its maximum at
\begin{align*}
\alpha_2 &= \arg \min_{\alpha} \left\| \Putildeperp \left( \tilde{\bz} - \alpha \tilde{\bv} \right) \right\|^2 
\\ &=
\arg \min_{\alpha} \left\| \Putildeperp \tilde{\bz} - \alpha \Putildeperp \tilde{\bv} \right\|^2
=
\left( \tilde{\bv}^{\dagger} \Putildeperp \tilde{\bv} \right)^{-1} \tilde{\bv}^{\dagger} \Putildeperp \tilde{\bz}
\end{align*}
and
\begin{eqnarray}
\label{eq:norma_pro_alpha}
\left\| \Putildeperp \left( \tilde{\bz} - \alpha \tilde{\bv} \right) \right\|^2 \!\!\!\!\!\! & =&
\left\| \Putildeperp \left( \tilde{\bz} - \left( \alpha -\alpha_2 \right)  \tilde{\bv} -\alpha_2 \tilde{\bv} \right) \right\|^2
\\ \nonumber &=&
\left\| \Putildeperp \tilde{\bz} -\alpha_2 \Putildeperp \tilde{\bv} 
 - \left( \alpha - \alpha_2 \right) \Putildeperp \tilde{\bv} 
\right\|^2
\\ \nonumber &=&
\left\| \left( \bI_N -  \Putildeperp \tilde{\bv} \left( \tilde{\bv}^{\dagger} \Putildeperp \tilde{\bv} \right)^{-1}  
\tilde{\bv}^{\dagger} \Putildeperp \right) \right.
\\ \nonumber &\times& \left. \Putildeperp \tilde{\bz} 
 - \left( \alpha -\alpha_2 \right) \Putildeperp \tilde{\bv} 
\right\|^2
\\ \nonumber &=&
\left\| \Pcomplex \Putildeperp \tilde{\bz} 
 - \left( \alpha -\alpha_2 \right) \Putildeperp \tilde{\bv} 
\right\|^2
\\ \nonumber &=& \tilde{\bz}^{\dagger}  \Putildeperp \Pcomplex \Putildeperp \tilde{\bz}
+ \left| \alpha -\alpha_2 \right|^2 \tilde{\bv}^{\dagger} \Putildeperp \tilde{\bv}
\\ \nonumber &=&
\tilde{\bz}^{\dagger}  \Putildeperp \tilde{\bz} 
- \tilde{\bz}^{\dagger}  \Putildeperp \tilde{\bv} \left( \tilde{\bv}^{\dagger} \Putildeperp \tilde{\bv} \right)^{-1} 
\\ \nonumber &\times&  \tilde{\bv}^{\dagger} \Putildeperp  \tilde{\bz}
+ \left| \alpha -\alpha_2 \right|^2 \tilde{\bv}^{\dagger} \Putildeperp \tilde{\bv}
\\ \nonumber &\geq&
\tilde{\bz}^{\dagger}  \Putildeperp \tilde{\bz} 
- \tilde{\bz}^{\dagger}  \Putildeperp \tilde{\bv}\! \left(\! \tilde{\bv}^{\dagger} \Putildeperp \tilde{\bv}\! \right)^{\!-1}\!\!\!  \tilde{\bv}^{\dagger} \Putildeperp  \tilde{\bz}.
\end{eqnarray}
Thus, $\forall y \geq \tilde{\bz}^{\dagger}  \Putildeperp \tilde{\bz} 
- \tilde{\bz}^{\dagger}  \Putildeperp \tilde{\bv} \left( \tilde{\bv}^{\dagger} \Putildeperp \tilde{\bv} \right)^{-1}  \tilde{\bv}^{\dagger} \Putildeperp  \tilde{\bz}$, the equation
$
\left\| \Putildeperp \left( \tilde{\bz} - \alpha \tilde{\bv} \right) \right\|^2 = y
$
is tantamount to 
$$
\left| \alpha - \alpha_2 \right|^2 \tilde{\bv}^{\dagger} \Putildeperp \tilde{\bv}=
y-
\tilde{\bz}^{\dagger}  \Putildeperp \tilde{\bz} 
+ \tilde{\bz}^{\dagger}  \Putildeperp \tilde{\bv} \left( \tilde{\bv}^{\dagger} \Putildeperp \tilde{\bv} \right)^{\!-1} \!\!\! \tilde{\bv}^{\dagger} \Putildeperp  \tilde{\bz}
$$
and is a circle centered at $\alpha_2$ of proper radius.
\medskip

Obviously, for $\alpha_1=\alpha_2$ the maximum 
of $l(b,\alpha)$, given $b$, is attained at $\alpha_1$.
Assuming instead $\alpha_1 \neq \alpha_2$,
we can prove that
the values of $\alpha$ maximizing $l(b,\alpha)$ belong
to the segment whose endpoints are $\alpha_1$ and $\alpha_2$, indicated hereafter 
as ${\cal S}_{\alpha_1\alpha_2}$.

\begin{figure}
\centering
\includegraphics[width=6cm]{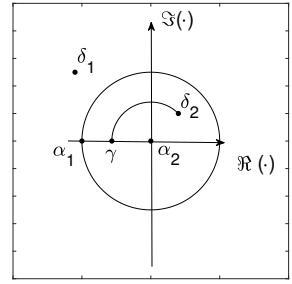}
\caption{Illustration of the procedure
to find points of
the ``segment $\alpha_1-\alpha_2$''
that
upperbound $l(b,\alpha)$. Without loss of generality $\alpha_2=0$ and $\alpha_1<0$.}
\label{fig:proof}
\end{figure}

To this end, we show that $\forall \alpha \in \C\setminus {\cal S}_{\alpha_1\alpha_2}$ there exists a 
point $\overline{\alpha}\in{\cal S}_{\alpha_1\alpha_2}$
such that
\begin{equation}
l(b,\alpha) < l(b,\overline{\alpha}).
\label{eq:upperbound_l}
\end{equation}
In fact, if
$\alpha \in \C$ does not belong to the circle centered at $\alpha_2$ of radius $|\alpha_1 - \alpha_2|$
as, for instance, $\delta_1$ in Fig.~\ref{fig:proof}, we can choose $\overline{\alpha}=\alpha_1$ that has a smaller distance from $\alpha_2$ than $\delta_1$,
thus implying
$
\left\| \Putildeperp \left( \tilde{\bz} - \delta_1 \tilde{\bv} \right) \right\|^2
>
\left\| \Putildeperp \left( \tilde{\bz} - \alpha_1 \tilde{\bv} \right) \right\|^2,
$
$
\left\|  \tilde{\bz} - \delta_1 \tilde{\bv} \right\|^2
 >
 \left\|  \tilde{\bz} - \alpha_1 \tilde{\bv} \right\|^2
$
and, eventually
$
l(b,\delta_1) < l(b,\alpha_1).
$

If, instead, $\alpha \in \C\setminus {\cal S}_{\alpha_1\alpha_2}$
belongs to the circle centered at $\alpha_2$ of radius $|\alpha_1 - \alpha_2|$
as, for instance, $\delta_2$ in Fig.~\ref{fig:proof},
we replace it with $\gamma$, i.e., $\overline{\alpha}=\gamma$, that has the same distance of $\delta_2$ from 
$\alpha_2$
and a smaller distance from $\alpha_1$, thus implying
$
\left\| \Putildeperp \left( \tilde{\bz} - \delta_2 \tilde{\bv} \right) \right\|^2
=
\left\| \Putildeperp \left( \tilde{\bz} - \gamma \tilde{\bv} \right) \right\|^2,
$
$
\left\|  \tilde{\bz} - \delta_2 \tilde{\bv} \right\|^2
 >
 \left\|  \tilde{\bz} - \gamma \tilde{\bv} \right\|^2
$
and, eventually
$
l(b,\delta_2) < l(b,\gamma).
$

\section*{Appendix B}
\renewcommand{\theequation}{B.\arabic{equation}}

\section*{
Proof of proposition~\ref{eq:GLRT-A}}

The derivative of $f(\nu)$ is given by
\begin{eqnarray*}
\frac{\mathrm{d}}{\mathrm{d}\nu}f(\nu) &=&
\frac{N}{K+1} \left(1+\nu\right)^{\frac{N}{K+1}-1}
\left( 1+ \frac{a}{1 + \nu} \right)
\\ &-&\left(1+\nu\right)^{\frac{N}{K+1}}\frac{a}{\left(1 + \nu \right)^2}.
\end{eqnarray*}
Thus,  $\frac{\mathrm{d}}{\mathrm{d}\nu}f(\nu) < 0$ if and only if
$
\frac{N}{K+1} \left(1+\nu +a \right)-a < 0
$
or, equivalently,
$
\nu < \left( \frac{K+1}{N} -1 \right) a-1.
$
It follows that the function $f(\nu)$ attains its minimum over $[0,+\infty)$
at  
$
\widehat{\nu}= 
\left( \frac{K+1}{N} -1 \right) a-1
$
if such a value is positive
and at $0$ otherwise.

\section*{
Characterization of detectors~(\ref{eq:1S-GLRT-5C})
and (\ref{eq:parametric_detector}).}

First we compute the $P_{fa}$ of the parametric detector (\ref{eq:parametric_detector}) and, as a special case, that of
the GLRT~(\ref{eq:1S-GLRT-5C}).
To this end, let
\be
\tilde{t}_K=\frac{t_K}{1-t_K}=
\frac{
\displaystyle{ \frac{|\bz^{\dagger} \bS^{-1} \bv|^{2}}{\bv^{\dagger} \bS^{-1} \bv} }
}{
1+\bz^{\dagger} \bS^{-1} \bz  -
\displaystyle{ \frac{|\bz^{\dagger} \bS^{-1} \bv|^{2}}{\bv^{\dagger} \bS^{-1} \bv}}
}
\label{eq:tilde_t_K}
\ee
where $t_K=\frac{|\bz^{\dagger} \bS^{-1} \bv|^{2}}{\bv^{\dagger} \bS^{-1} \bv \left( 1+\bz^{\dagger} \bS^{-1} \bz \right)}$ is Kelly's statistic 
and
\be
b= 
\frac{1}{
1+\bz^{\dagger} \bS^{-1} \bz  -
\displaystyle{ \frac{|\bz^{\dagger} \bS^{-1} \bv|^{2}}{\bv^{\dagger} \bS^{-1} \bv}}
}.
\label{eq:b}
\ee
It turns out that
$
1+\left\|  \tilde{\bz}  \right\|^2 = 1+\bz^{\dagger} \bS^{-1} \bz= \left(1+\tilde{t}_K\right)/b
$
and
$
\left\| \Pvtildeperp \tilde{\bz} \right\|^2 = \bz^{\dagger} \bS^{-1} \bz - \frac{|\bz^{\dagger} \bS^{-1} \bv|^{2}}{\bv^{\dagger} \bS^{-1} \bv} = \frac{1}{b}-1
$
and, hence,
the decision statistic of the  detector 
(\ref{eq:parametric_detector})
can be re-written as
\begin{equation}
\Lambda_\epsilon(\bz, \bS) =
\left\{
\begin{array}{ll}
\frac{
\frac{1+\tilde{t}_K}{b} \left( 1-\frac{1}{\zeta_{\epsilon}} \right)
}{\left[ \left(\zeta_{\epsilon} -1 \right)
\left( \frac{1}{b}-1\right) \right]^{\frac{1}{\zeta_{\epsilon}}}
}, &
b < 1-\frac{1}{\zeta_{\epsilon}} \\
1+\tilde{t}_K, & \mbox{otherwise.}
\end{array}
\right. 
\label{eq:GLRT_Sigma=C}
\end{equation}

Under the noise-only hypothesis
the random variable (RV) $\tilde{t}_K$, given by equation (\ref{eq:tilde_t_K}), is distributed
according to a complex central F-distribution
with $1$ and $K-N+1$ (complex) degrees of freedom and
it is independent of $b$, 
given by equation (\ref{eq:b}),
which, in
turn, obeys a complex central beta distribution with $K-N+2$ and $N-1$ (complex) degrees
of freedom~\cite{Kelly,Kelly_techrep,BOR-MorganClaypool}. In symbols, we write
$\tilde{t}_K \sim {\cal CF}_{1,K-N+1}$ and $b \sim {\cal C\beta}_{K-N+2,N-1}$.
It is thus apparent that $P_{fa}$ is independent of $\bC$ and, hence,  the detector possesses the 
CFAR property.

Moreover, we can compute $P_{fa}$ as follows
\[
P_{fa}=\int_{0}^1 \textrm{P} [\Lambda_\epsilon(\bz, \bS) >\eta | b=x, H_0] f_{b} (x) \, \mathrm{d} x
\]
where $f_{b} (\cdot)$ denotes the PDF of the RV $b$ \cite{BOR-MorganClaypool}, i.e.,
$$
f_{b} (x)= 
\frac{1}{B(K+2-N,N-1)}
x^{K-N+1}(1-x)^{N-2}
$$
with $0 \leq x \leq 1$, while
\begin{eqnarray*}
p(x) \!\!\! &=& \textrm{P} [\Lambda_\epsilon(\bz, \bS) > \eta | b=x, H_0]
\\ &=&
\left\{
\begin{array}{ll}
1-F_{\tilde{t}_K} \left(y_\epsilon(x)\right), & 0 \leq x \leq 1 - \frac{1}{\zeta_{\epsilon}} 
\\
1-F_{\tilde{t}_K}(\eta-1), &1 - \frac{1}{\zeta_{\epsilon}} \leq x \leq1
\end{array}
\right.
\end{eqnarray*}
with \cite[formula (A.7)]{BOR-MorganClaypool}
$$
1-F_{\tilde{t}_K}(y_\epsilon)=
\left\{
\begin{array}{ll}
\frac{1}{(1+y_\epsilon)^{K-N+1}}, & y_\epsilon \geq 0 \\
1, & y_\epsilon < 0
\end{array}
\right.
$$
and
$
y_\epsilon(x)= \eta x \left( \left(\zeta_{\epsilon}-1\right) \frac{1-x}{x} \right)^{\frac{1}{\zeta_{\epsilon}}}
\frac{\zeta_{\epsilon}}{\zeta_{\epsilon}-1}-1.
$
To compute the $P_{fa}$, we  observe that
the derivative of 
$y_\epsilon(x)$
is given by
$$
\frac{\mathrm{d} y_\epsilon(x)}{\mathrm{d}x} \!= \eta \left( \left( \zeta_{\epsilon}-1 \right) \frac{1-x}{x} \right)^{\!\frac{1}{\zeta_{\epsilon}}-1} \!\!\!
\left( \zeta_{\epsilon} \frac{1-x}{x} - \frac{1}{x}  \right)
$$
and it is positive for $0 < x < 1-\frac{1}{\zeta_{\epsilon}}$.
Moreover, 
$
\lim_{x \rightarrow 0^{+}} y_\epsilon(x) =-1
$
and 
$
y_\epsilon \left( 1-\frac{1}{\zeta_{\epsilon}} \right)= \eta-1.
$
It turns out that $P_{fa}=1$ for $\eta \leq 1$.
If, instead, $\eta >1$, denoting by $\overline{x}_\epsilon(\eta)$ that number in $(0, 1-\frac{1}{\zeta_{\epsilon}})$ such that $y_\epsilon(\overline{x})=0$,
it follows that
\begin{align*}
P_{fa} &= \int_{0}^{\overline{x}_\epsilon(\eta)} f_b(x) \, \mathrm{d}x  +
\int_{\overline{x}_\epsilon(\eta)}^{1-\frac{1}{\zeta_{\epsilon}}}
\frac{1}{
\left[ 1+y_\epsilon(x) \right]^{K+1-N} 
} f_b(x) \, \mathrm{d}x
\\  &+
\frac{1}{\eta^{K+1-N}} \int_{1-\frac{1}{\zeta_{\epsilon}}}^1 f_b(x) \, \mathrm{d}x 
\end{align*}
i.e., introducing $
c= \frac{1}{B(K+2-N,N-1)} = \frac{K!}{(K+1-N)!(N-2)!}
$,
\begin{align*}
P_{fa} &=
c \int_{0}^{\overline{x}_\epsilon(\eta)} x^{K-N+1} (1-x)^{N-2} \, \mathrm{d}x 
\\ &+
\frac{a_\epsilon c}{\eta^{K+1-N}} \!\!
\int_{\overline{x}_\epsilon(\eta)}^{1-\frac{1}{\zeta_{\epsilon}}}
x^{\frac{K+1-N}{\zeta_{\epsilon}}} (1-x)^{N-2-\frac{K+1-N}{\zeta_{\epsilon}}} \, \mathrm{d}x
\\ &+
\frac{c}{\eta^{K+1-N}} 
\int_{1-\frac{1}{\zeta_{\epsilon}}}^{1} x^{K-N+1} (1-x)^{N-2} \, \mathrm{d}x
\end{align*} 
where
$
a_\epsilon=\frac{\left(\frac{\zeta_{\epsilon}-1}{\zeta_{\epsilon}} \right)^{K+1-N}}{
\left(\zeta_{\epsilon}-1 \right)^{\frac{K+1-N}{\zeta_{\epsilon}} }
}
$.
The $P_{fa}$ of (\ref{eq:parametric_detector}) is thus given by
\begin{eqnarray}
P_{fa} &=&
\nonumber
c B_{\overline{x}_\epsilon(\eta)}(K-N+2,N-1) + \frac{a_\epsilon c}{\eta^{K+1-N}} 
\\ \nonumber &\times&
\left[
B_{1-\frac{1}{\zeta_{\epsilon}}} \left( \frac{K+1-N}{\zeta_{\epsilon}}+1, N-1-\frac{K+1-N}{\zeta_{\epsilon}} \right) \right.
\\ \nonumber &-&
\left. B_{\overline{x}_\epsilon (\eta)} \left( \frac{K+1-N}{\zeta_{\epsilon}}+1, N-1-\frac{K+1-N}{\zeta_{\epsilon}} \right) 
\right]
\\ \nonumber &+&
\frac{c}{\eta^{K+1-N}} 
\Big[ B \left(K-N+2,N-1\right) 
\\ &-&  B_{1-\frac{1}{\zeta_{\epsilon}}}  \left(K-N+2,N-1\right)
\Big]
\label{eq:PFA-proposed}
\end{eqnarray} 
where 
$B(\cdot,\cdot)$ is the Eulerian beta function
while
$
B_u(\nu,\mu)= \int_0^u t^{\nu-1} (1-t)^{\mu-1} \, \mathrm{d}t
$
is the incomplete beta function \cite{Ryzhik}.

As a special case, i.e., $\epsilon=0$ and, hence, $\zeta_{\epsilon}=\zeta$, we obtain the $P_{fa}$ of the GLRT~(\ref{eq:1S-GLRT-5C}). 
\medskip

On the other hand, if we suppose that under the $H_1$ hypothesis
the actual useful signal is deterministic, but with a steering vector $\bp$ different
from the nominal one $\bv$, i.e.,
\[
\bor=\alpha\bp + \bn, \quad  \bn \sim {\cal CN}_N(\bzero,\bC),
\]
$\tilde{t}_K$, given $b$, is ruled by
a complex noncentral F-distribution with
$1$ and $K-N+1$ (complex) degrees of freedom and noncentrality
parameter $\delta$, with \cite{Kelly89,BOR-MorganClaypool,KellyTR2}
\[
\delta^2=\mbox{SNR}  \cdot b  \cos^2\theta,
\]
where
\be
\mbox{SNR}=|\alpha|^2 \ \bp^\dag \bC^{-1}\bp
\label{eqn:defSNR}
\ee
and
\be
\cos^2\theta=\frac{|\bp^\dag \bC^{-1}\bv|^2}{(\bv^\dag\bC^{-1}\bv)(\bp^\dag
\bC^{-1}\bp)}.
\label{eqn:defcostheta}
\ee
In addition, the RV $b$
obeys the complex noncentral beta distribution
with $K-N+2$ and $N-1$ (complex) degrees of freedom and
noncentrality parameter $\delta_b$, with
\[
\delta_{b}^2=\mbox{SNR} \cdot \sin^2\theta.
\]
In symbols, we write
$\tilde{t}_K \sim {\cal CF}_{1,K-N+1}(\delta)$ and $b \sim {\cal C\beta}_{K-N+2,N-1}(\delta_b)$.

In the special case of perfect match between $\bp$ and $\bv$, then
$\delta^2 =|\alpha|^2 \ \bv^\dag \bC^{-1}\bv \cdot b$ and $\delta_{b}^2=0$, i.e., $b \sim {\cal C \beta}_{K-N+2,N-1}$
and $\tilde{t}_K \sim {\cal CF}_{1,K-N+1}(\delta)$ given $b$.

The above characterization highlights that performance of detectors~(\ref{eq:1S-GLRT-5C}) and (\ref{eq:parametric_detector})
can be expressed in terms of $P_d$ vs  
$\mbox{SNR}$ (equation~(\ref{eqn:defSNR})) given $\cos^2 \theta$ (equation~(\ref{eqn:defcostheta}))
and $P_{fa}$.
In principle, such characterization could be exploited to 
compute $P_d$, paralleling the computation of $P_{fa}$; however, analytical forms for $P_d$
are less useful since Monte Carlo simulation is not very time-consuming.
As a matter of fact, we use Monte Carlo simulation to compute $P_d$ vs SNR, 
expressed by equation~(\ref{eqn:defSNR}),
given $\cos^2 \theta$ and $P_{fa}$.

\section*{Appendix C}
\renewcommand{\theequation}{C.\arabic{equation}}

\section*{The two-step GLRT-based detector with
$\bSigma=\bv\bv^\dag$ is the AMF}

We focus
on the binary hypotheses testing problem
$$
\left\{
\begin{array}{ll}
H_0:  & \bz \sim {\cal CN}_N(\bzero, \bC) \\
H_1:  & \bz \sim {\cal CN}_N(\alpha  \bv, \nu \bu \bu^{\dagger} + \bC) 
\end{array}
\right.
$$
where 
$\nu \geq 0$ and $\alpha \in \C$
are unknown quantities while the vectors $\bu, \bv \in \C^{N \times 1}$ and the Hermitian positive definite matrix $\bC$
are known. 
The corresponding GLRT  is given by
\be
\Lambda(\bz, \bC)=
\frac{
\displaystyle{\max_{\nu \geq 0} \max_{\alpha \in \C} f_1( \bz | \nu, 
\alpha)
}}{
\displaystyle{f_0( \bz )
}}
\test \eta
\label{eq:2S-GLRT-uuH}
\ee
where 
$$
f_1( \bz | \nu, \alpha) 
=
\frac{1}{\pi^{N}} \frac{1}{(1+\nu \bu^\dag \bC^{-1}\bu) \det(\bC)} \ l(\nu,\alpha)
$$
is the PDF of $\bz$
under $H_1$ 
with
$$
l(\nu,\alpha)=
e^{- \displaystyle{
\left( \bz - \alpha \bv \right)^{\dagger}\bC^{-1}\left( \bz - \alpha \bv \right) + \nu \frac{\left|\bu^\dag \bC^{-1}  \left(\bz - \alpha \bv\right)\right|^2}{1+\nu \bu^\dag \bC^{-1}\bu}}}
$$
while $f_0( \bz )$, the PDF of $\bz$ under $H_0$,
is given by 
\begin{equation*}
f_0( \bz ) = \frac{1}{\pi^{N} \det(\bC)} 
e^{-  \bz^{\dagger}\bC^{-1}\bz } .
\end{equation*}

To maximize $l(\nu,\alpha)$
over
$\alpha$ we rewrite its exponent as
\begin{eqnarray*}
&& - \left( \bz - \alpha \bv \right)^{\dagger}\bC^{-1}\left( \bz - \alpha \bv \right) + \nu \frac{\left|\bu^\dag \bC^{-1}  \left(\bz - \alpha \bv\right)\right|^2}{1+\nu \bu^\dag \bC^{-1}\bu} 
\\ &=&
- \left\| \tilde{\bz} - \alpha \tilde{\bv} \right\|^2 + \nu \frac{\left|\tilde{\bu}^\dag 
\left(\tilde{\bz} - \alpha \tilde{\bv}\right)\right|^2}{1+\nu \left\| \tilde{\bu} \right\|^2}
\end{eqnarray*}
where 
$\tilde{\bz} = \bC^{-1/2} \bz$, $\tilde{\bv}= \bC^{-1/2} \bv$, and $\tilde{\bu}= \bC^{-1/2} \bu$. Using
$$
\left\| \tilde{\bz} - \alpha \tilde{\bv} \right\|^2=
\left\| \Putilde \left( \tilde{\bz} - \alpha \tilde{\bv} \right) \right\|^2
+
\left\| \Putildeperp \left( \tilde{\bz} - \alpha \tilde{\bv} \right) \right\|^2
$$
and
$$
\left| \tilde{\bu}^{\dagger} \left( \tilde{\bz} - \alpha \tilde{\bv} \right) \right|^2
=\left\| \tilde{\bu} \right\|^2 \left\| \Putilde \left( \tilde{\bz} - \alpha \tilde{\bv} \right) \right\|^2
$$
we obtain
$$
l(\nu,\alpha)=
e^{\displaystyle{
- \frac{\left\| \tilde{\bz} - \alpha \tilde{\bv} \right\|^2 + \nu 
\left\| \tilde{\bu} \right\|^2 \left\| \Putildeperp \left( \tilde{\bz} - \alpha \tilde{\bv} \right) \right\|^2
}{1+\nu \left\| \tilde{\bu} \right\|^2}
}}.
$$
Now letting $\bu=\bv$ and, hence, $\tilde{\bu}=\tilde{\bv}$ we have
an expression
that can be easily maximized with respect to $\alpha$
obtaining
$$
\max_{ \alpha \in \C} l(\nu,\alpha)=
e^{ -\left\| \Pvtildeperp \tilde{\bz} \right\|^2}
$$
and hence
$$
\max_{ \alpha \in \C} f_1( \bz | \nu, \alpha) =
\frac{1}{\pi^{N}} \frac{1}{(1+\nu \bv^\dag \bC^{-1}\bv) \det(\bC)} 
e^{ -\left\| \Pvtildeperp \tilde{\bz} \right\|^2}.
$$
It follows that
$$
\max_{\nu \geq 0} \max_{\alpha \in \C} f_1( \bz | \nu, \alpha) =
\frac{1}{\pi^{N}} \frac{1}{ \det(\bC)} 
e^{ -\left\| \Pvtildeperp \tilde{\bz} \right\|^2}
$$
and
\be
\Lambda(\bz, \bC)=
e^{\left\| \Pvtilde \tilde{\bz} \right\|^2}
=
e^{ \frac{\left| \bz^{\dagger} \bC^{-1} \bv \right|^2}{\bv^{\dagger} \bC^{-1} \bv } }
\test \eta.
\ee
The corresponding adaptive detector, obtained by replacing $\bC$ with the sample covariance matrix based on the secondary data, is obviously equivalent to the AMF.

\begin{IEEEbiography}
 [{\includegraphics[width=1in,height=1.25in,clip,keepaspectratio]{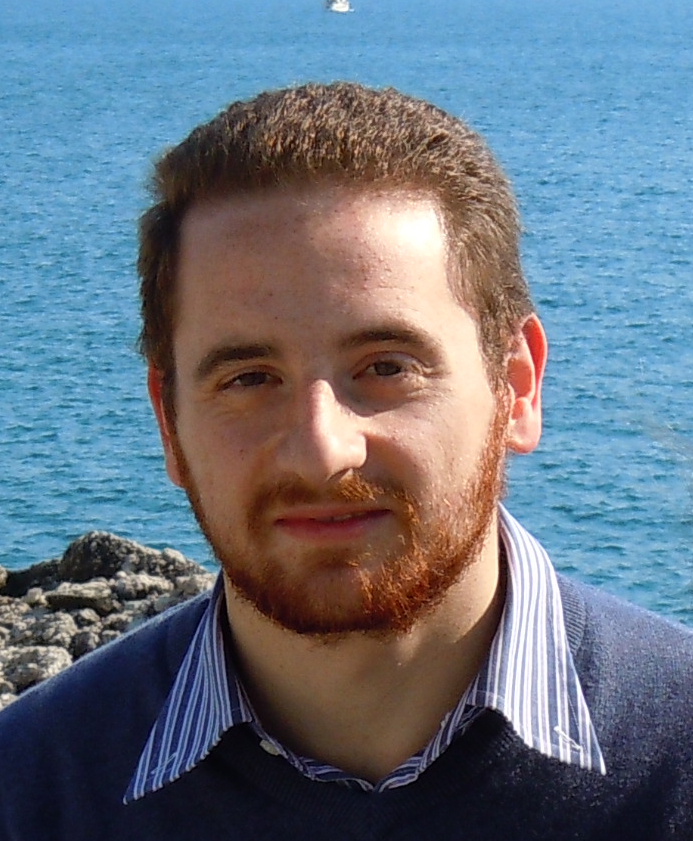}}]
{Angelo Coluccia} (M'13--SM'16) received the Eng. degree in Telecommunication Engineering (summa cum laude) in 2007 and the PhD degree in Information Engineering in 2011, both from the University of Salento (Lecce, Italy). He is completing a tenure-track for Associate Professor at the Dipartimento di Ingegneria dell'Innovazione, University of Salento, where he teaches Telecommunication Systems. Since 2008 he has been engaged in European research projects on different topics, including small-drone detection, tracking, and data fusion. He has been researcher at Forschungszentrum Telekommunikation Wien (Vienna, Austria), working on mobile Internet traffic modeling and anomaly detection. He has held a visiting position at the Department of Electronics, Optronics, and Signals (DEOS) of the Institut Sup\'erieur de l'A\'eronautique et de l'Espace (ISAE-Supa\'ero, Toulouse, France), in the Signal, Communication, Antennas and Navigation (SCAN) Group. His research interests are multi-sensor and multi-agent statistical signal processing for detection, estimation, learning, and localization problems. Relevant application fields are radar, wireless networks (including 5G), intelligent cyber-physical systems, and emerging network contexts (including smart devices and social networks). He is Member of the Special Area Team in Signal Processing for Multisensor Systems of EURASIP.
\end{IEEEbiography}

\begin{IEEEbiography}
 [{\includegraphics[width=1in,height=1.25in,clip,keepaspectratio]{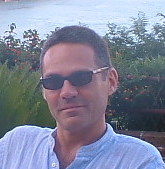}}]
{Giuseppe Ricci} 
(M'01--SM'10) was born in Naples, Italy, on February 15, 1964. He received the Dr. degree and the Ph.D. degree, both in Electronic Engineering, from the University of Naples ``Federico II'' in 1990 and 1994, respectively. Since 1995 he has been with the University of Salento (formerly University of Lecce) first as an Assistant Professor of Telecommunications and, since 2002, as a Professor. His research interests are in the field of statistical signal processing with emphasis on radar processing, localization algorithms, and CDMA systems. More precisely, he has focused on high-resolution radar clutter modeling, detection of radar signals in Gaussian and non-Gaussian disturbance, oil spill detection from SAR data, track-before-detect algorithms fed by space-time radar data, localization in wireless sensor networks, multiuser detection in overlay CDMA systems, and blind multiuser detection. He has held visiting positions at the University of Colorado at Boulder (CO, USA) in 1997-1998 and in April/May 2001, at the Colorado State University (CO, USA) in July/September 2003, March 2005, September 2009, and March 2011, at Ensica (Toulouse, France) in March 2006, and at the University of Connecticut (Storrs CT, USA) in September 2008. 
\end{IEEEbiography}

\begin{IEEEbiography}
 [{\includegraphics[width=1in,height=1.25in,clip,keepaspectratio]{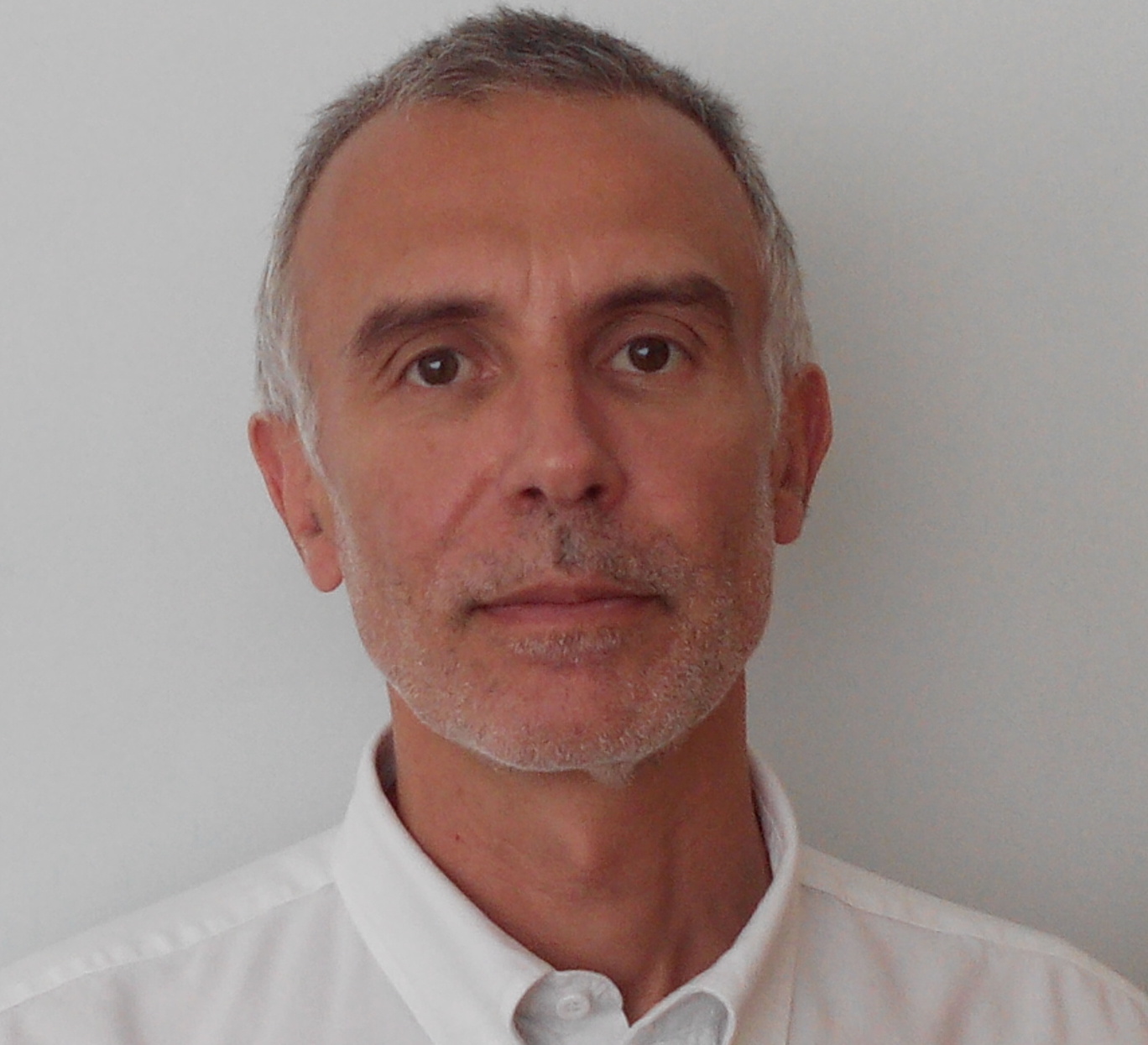}}]
{Olivier Besson} 
received the PhD degree in signal processing from Institut National Polytechnique Toulouse in 1992. He is currently a Professor with Institut Sup\'erieur de l'A\'eronautique et de l'Espace (ISAE-SUPAERO), Toulouse, France. His research interests are in statistical array processing, multivariate analysis, adaptive detection and estimation.
\end{IEEEbiography}


\begin{thebibliography}{99}

\bibitem{Ward}
J. Ward, ``Space-Time Adaptive Processing for Airborne Radar,''
Lincoln Laboratory, MIT, Lexington, MA, Tech. Rep. No. 1015, Dec. 1994.

\bibitem{Klemm-STAP}
R. Klemm, {\em Principles of Space-Time Adaptive Processing}, IEE
Radar, Sonar, Navigation and Avionics Series 12, 2002.


\bibitem{Kelly}
E. J. Kelly, ``An Adaptive Detection Algorithm,'' {\em IEEE Trans. Aerosp. and Electron. Syst.}, Vol. 22, No.~2, pp.~115-127, Mar.~1986.


\bibitem{Kelly89}
E. J. Kelly, ``Performance of an Adaptive
Detection Algorithm; Rejection
of Unwanted Signals,''
{\em IEEE Trans. Aerosp. and Electron. Syst.},
Vol. 25, No.~2, pp.~122-133, Mar.~1989.


\bibitem{Kelly-Nitzberg}
F.~C.~Robey, D.~L.~Fuhrman, E.~J.~Kelly, and R.~Nitzberg,
``A CFAR Adaptive Matched Filter Detector,''
{\em IEEE Trans. Aerosp. and Electron. Syst.},
Vol.~29, No.~1, pp.~208-216, Jan.~1992.


\bibitem{ACE}
S. Kraut and L. L. Scharf, ``The CFAR Adaptive Subspace Detector is
a Scale-Invariant GLRT,'' {\em IEEE Trans. Signal Process.}, Vol.~47, No. 9, pp.~2538-2541, Sept. 1999.

\bibitem{Pulsone-Rader}
N. B. Pulsone and C. M. Rader, ``Adaptive Beamformer Orthogonal
Rejection Test,'' {\em IEEE Trans. Signal Process.}, Vol.~49,
No.~3, pp.~521-529, Mar. 2001.

\bibitem{Fabrizio-Farina}
G. A. Fabrizio, A. Farina, and M. D. Turley, ``Spatial Adaptive Subspace Detection in OTH Radar,''
{\em IEEE Trans. Aerosp. and Electron. Syst.},
Vol. 39, No. 4, pp. 1407-1427, Oct. 2003.

\bibitem{W-ABORT}
F.~Bandiera, O.~Besson, and G.~Ricci, ``An ABORT-Like Detector
With Improved Mismatched Signals Rejection Capabilities,''
{\em IEEE Trans. Signal Process.}, Vol. 56, No. 1, pp. 14-25, Jan. 2008.


\bibitem{Kalson}
S. Z. Kalson, ``An Adaptive Array Detector with Mismatched Signal Rejection,''
{\em IEEE Trans. Aerosp. Electron. Syst.}, Vol. 28, No. 1, pp. 195-207, Jan. 1992.

\bibitem{KWA} 
F. Bandiera, D. Orlando, and G. Ricci, ``One- and Two-Stage Tunable Receivers*'', \emph{IEEE Trans. Signal Process.}, Vol. 57, No. 8, Aug. 2009.

\bibitem{JunLiu1}
J. Liu, S. Zhou, W. Liu, J. Zheng, H. Liu, and J. Li, ``Tunable Adaptive Detection in Colocated MIMO Radar,'' \emph{IEEE Trans. Signal Process.}, Vol. 66, No. 4, pp. 1080-1092, Feb. 15, 2018.

\bibitem{JunLiu2}
J. Liu, H.-Y. Zhao, W. Liu, H. Li, and H. Liu, ``Adaptive Detection Using both the Test and Training Data for Disturbance Correlation Estimation,'' \emph{Signal Process.}, Vol. 137, pp. 309-318, 2017.

\bibitem{Coni-SOC}
A. De Maio, ``Robust Adaptive Radar Detection in the Presence of
Steering Vector Mismatches,'' {\em IEEE Trans. Aerosp. Electron. Syst.}, vol.
41, No. 4, pp. 1322-1337, Oct. 2005.

\bibitem{Besson1}
O. Besson, ``Detection of a Signal in Linear Subspace with
Bounded Mismatch,'' {\em IEEE Trans. Aerosp. Electron. Syst.}, Vol.
42, No. 3, pp. 1131-1139, Jul. 2006.

\bibitem{Chenpeng}
C. Hao, B. Liu, and L. Cai, 
``Performance Analysis of a Two-Stage Rao Detector,''
{\em Signal Process.}, Vol. 91, pp. 2141-2146, 2011.

\bibitem{JunLiu3}
J. Liu, W. Liu, B. Chen, H. Liu, H. Li, and C. Hao, ``Modified Rao Test for Multichannel Adaptive Signal Detection,'' {\em IEEE Trans. Signal Process.}, Vol. 64, No. 3, pp. 714-725, Feb. 1, 2016.

\bibitem{Liu}
J. Liu,  K. Li, X. Zhang, M. Liu, W. Liu,
``A Weighted Detector for Mismatched Subspace Signals,''
{\em Signal Process.}, Vol. 140, pp. 110-115, 2017.

\bibitem{Duan}
K. Duan, M. Liu, H. Dai, F. Xu, and W. Liu,
``A Two-stage Detector for Mismatched Subspace Signals,''
{\em IEEE Geosci. Remote Sens. Lett.}, Vol. 14, No. 12, pp. 2270-2274, Dec. 2017.

\bibitem{Orlando1}
D. Ciuonzo, A. De Maio, and D. Orlando,
``A Unifying Framework for Adaptive Radar Detection in Homogeneous plus Structured Interference - Part II: Detectors Design,'' {\em IEEE Trans. Signal Process.}, Vol. 64, No. 11, pp. 2907-2919, Jun. 1, 2016.

\bibitem{Orlando2}
D. Ciuonzo, A. De Maio, and D. Orlando,
``On the Statistical Invariance for Adaptive Radar Detection in Partially Homogeneous Disturbance plus Structured Interference,'' {\em IEEE Trans. Signal Process.}, Vol. 65, No. 5, pp. 1222-1234, Mar. 1, 2017.

\bibitem{CAMSAP}
O. Besson, E. Chaumette, and F. Vincent,
``Adaptive Detection of a Gaussian Signal in Gaussian Noise,''
{\em The Sixth IEEE International Workshop on Computational Advances in Multi-Sensor Adaptive 
Processing CAMSAP 2015}, Canc\'un, Mexico, 13-16 Dec. 2015.

\bibitem{Besson_collaboration}
O. Besson, A. Coluccia, E. Chaumette, G. Ricci, and F. Vincent,
``Generalized Likelihood Ratio Test for Detection of Gaussian Rank-One Signals 
in Gaussian Noise with Unknown Statistics,''
{\em IEEE Trans. Signal Process.}, Vol. 65, No. 4, Feb. 15, 2017.



\bibitem{CISS2018}
A. Coluccia and G.~Ricci, 
``A Random-Signal Approach to Robust Radar Detection,''
{\em 52nd Annual Conference on Information Sciences and Systems (CISS)},
Princeton, NJ, USA, 21-23 Mar. 2018.

\bibitem{Horn-Johnson}
R. A. Horn and C. R. Johnson, {\em Matrix Analysis}, Cambridge University Press, 1985.

\bibitem{Kelly_techrep}
E. J. Kelly and K. Forsythe, ``Adaptive Detection and Parameter
Estimation for Multidimensional Signal Models,'' Lincoln Laboratory, MIT,
Lexington, MA, Tech. Rep. No. 848, Apr. 19, 1989.

\bibitem{BOR-MorganClaypool}
F. Bandiera, D. Orlando, and G. Ricci,
``Advanced Radar Detection Schemes Under Mismatched Signal Models,''
{\em Synthesis Lectures on Signal Processing No. 8, Morgan \& Claypool Publishers},
2009.

\bibitem{Ryzhik}
L. S. Gradshteyn, I. M. Ryzhik, {\em Table of Integrals, Series, and Products},
Academic Press, 6th ed., 2000.

\bibitem{KellyTR2} 
E. J. Kelly, ``Adaptive Detection in Non-Stationary Interference-Part III,''
Lincoln Laboratory, MIT, Lexington, MA, Tech. Rep. No. 761, Aug. 1987.

\bibitem{SP2020}
A. Coluccia, A. Fascista, and G. Ricci, ``A Novel Approach to Robust Radar Detection of Range-Spread Targets,''
Signal Processing, Vol. 166, Jan. 2020. Available online: https://doi.org/10.1016/j.sigpro.2019.07.016

\end{thebibliography}
\end{document}